\documentstyle[epsf,a4wide]{article}

\advance\hoffset by -.5 cm
\begin{document}

\begin{tabbing}
\hskip 14 cm \= {MRAO-1917}\\
\> DAMTP-96-26\\
\>Imperial/TP/95-96/47\\
\> CfPA-96-th-11\\
\> Submitted to {\em PRD}\\
\> May 1996 \\
\end{tabbing}

\openup1\jot

\begin{center}
{\Huge\bf    
The structure of Doppler peaks induced by active perturbations}
\vskip 1.2cm
{\large \bf Joao Magueijo $^1$, Andreas Albrecht$^2$, Pedro Ferreira$^3$,
David Coulson$^4$}\\
$^{(1)}$Department of Applied Mathematics and Theoretical
Physics\\University of Cambridge, Cambridge CB3 9EW, UK\\
and\\ Mullard Radio Astronomy Observatory,
Cavendish Laboratory\\ Madingley Road, Cambridge, CB3 0HE, UK\\
{$^{(2)}$Blackett Laboratory, Imperial College, Prince Consort Road\\
 London SW7 2BZ  U.K.\\
$^{(3)}$Center for Particle Astrophysics, University of
California Berkeley,  CA 94720-7304\\
$^{(4)}$D. Rittenhouse Laboratory, University of Pennsylvania\\
 Philadelphia,  PA, 19104, USA
}
\end{center}

\begin{flushleft}
\end{flushleft}

\begin{abstract}
\openup1\jot
We investigate how the qualitative structure of Doppler peaks 
in the angular power spectrum of the cosmic microwave anisotropy
is affected by basic assumptions going into theories of structure formation.
We define the  
concepts of  ``coherent'' and ``incoherent'' fluctuations,
and also of ``active'' and ``passive'' fluctuations. 
In these terms inflationary
fluctuations are passive and coherent  while topological defects
are active incoherent fluctuations. Causality and
scale invariance are shown to have different implementations in theories 
differing in the above senses. 
We then extend the
formalism of Hu and Sugiyama to treat models with cosmic
defects. Using this formalism we show 
that the existence or absence of secondary 
Doppler peaks and the rough placing of the primary peak 
are very sensitive to  the fundamental properties defined.
We claim therefore that even a rough measurement of the angular 
power spectrum $C_l$ shape at $100<l<1500$ ought to tell us which are  
the basic ingredients to be used in the right structure formation
theory. 
We also apply our formalism to cosmic string theories.
These are shown to fall into the class of active incoherent theories
for which one can robustly predict the absence of secondary Doppler peaks.
The placing of the cosmic strings' primary
peak is more uncertain, but should fall in $l\approx400-600$.
\end{abstract}

\thispagestyle{empty}
\pagebreak
\setcounter {page}{1}

\section{Introduction}
The cosmic microwave background (CMB)  promises to become 
one of the most successful bridges between theory 
and experiment in cosmology. As the body of experimental
data continues to grow\cite{exp}, theorists
are evaluating the impact of this data on the two major
paradigms for cosmic structure formation: inflation \cite{infl} and
topological defects \cite{defc}. 
The so-called Doppler peaks, in particular, have attracted
great  interest.
They consist of a system of oscillations, known to be 
present for most inflationary models, 
in the CMB angular power spectrum $C_l$ at   $100< l < 1500$. 

The Doppler peaks' height 
and position have been extensively studied in inflationary
scenarios \cite{BE}, a context in which they 
can be used to fix with some accuracy combinations 
of cosmological parameters \cite{crit}. Although there has been some
work on defect models \cite{bsb,b&r,reion},
progress on defect Doppler peak predictions has been slow
(see however \cite{neil,durrer,us}). 
In two recent letters \cite{us,usagain}, we have claimed
that regardless  of the remaining quantitative uncertainties,
one could expect dramatic qualitative differences between
defect and inflationary Doppler peaks.
In summary, we found two types of exotic behaviour.
Firstly we found that defect Doppler peaks may
be obtained from the inflationary ones by an additive shift in $l$.
The shift value is roughly proportional to the inverse of the defect scaling
coherence length. Only very large defects, 
on the verge of violating causality, apply a zero shift.
We also found that defects should always soften 
the secondary Doppler oscillations, the more so the larger the additive shift
they apply (the smaller their coherence length). Only ``zero-shift''
defects (placing the peaks on the inflationary positions)
induce a negligible softening. If the main peak is at the isocurvature
position one should not obtain more than an undulation. For a smaller
coherence length (main peak to the right of the isocurvature position)
the secondary peaks should be completely erased. These two types of phenomena
are unheard of in inflationary scenarios. Putting aside ``zero-shift''
defects, we then claimed that measuring
the qualitative features of the $C_l$ spectrum at $100<l<1500$ 
should decide between inflation and defects on fairly general grounds.
 
In this paper we explain in more detail the formalism outlined in
\cite{usagain}, and present our findings in a more systematic form.
As in \cite{usagain}
we focus on the basic assumptions of
inflationary and defect theories in order to  isolate  the 
contrasting properties which alone determine 
the broad Doppler peak features (Section~\ref{definfl}).
We come up with two concepts: the concept of active and passive perturbations,
and the concept of coherent and incoherent perturbations.   
In terms of these inflationary fluctuations are passive and coherent. 
Topological defects are active incoherent fluctuations. Causality and
scale invariance are shown to have different implementations in theories 
differing in the above senses. These two concepts then allow us
to write down a Hu and Sugiyama (H+S)
type of formalism \cite{HS} tailored for cosmic defects 
(Sections~\ref{hsreview} and \ref{hsdefect}).
A number of simplifications apply, but one also needs to extend the existing
work in two ways. Firstly, the way in which averages are taken has 
to be modified for incoherent perturbations. Two approximations may 
be useful: the coherent and the totally incoherent approximations.
The factors controlling the applicability of these approximations
are spelled out in Section~\ref{qualinc}. Secondly the H+S formalism
cannot account for radiation backreaction, that is, for the radiation
itself acting as a source for its own driving force. Causality, in defect
theories, requires the existence of a large scale nonignorable radiation
white noise-spectrum. This leads to a backreaction effect which can
never be neglected. In Section~\ref{causal} we show how to solve this
problem.

We then apply our formalism in two types of study. In 
Section~\ref{genqual} we perform a qualitative analysis of defect
Doppler peaks for a generic defect. 
We explain in more detail the
results concerning the position
of the primary peak and structure of secondary oscillations
published in \cite{usagain} and highlighted above.
We go beyond the coherent and totally incoherent approximations.
We also explicitly translate into $C_l$ spectra arguments
initially developed in terms of the photons' energy power spectrum
at last scattering.

On a different front we target cosmic strings (Section~\ref{cs}).
By using results from a cosmic string simulation we show that regardless of 
uncertainties one can safely predict that strings do not
exhibit secondary oscillations. We also argue that the totally
incoherent  
approximation is applicable to cosmic strings, and solve the full
H+S algorithm in this approximation. The cosmic string
(single) Doppler peak is shown to be in the region $l\approx 400-600$.
Our simulation and formalism uncertainties do not yet allow us to
make any prediction concerning the peak's height relative to the low
$l$ plateau.

\section{Review of the Hu and Sugiyama formalism}\label{hsreview}
We start by highlighting results in \cite{HS} of which we
shall make use. The CMB photons are described by
the brightness function $\Theta({\bf x},\mu, \eta)$, defined as
the brightness temperature of photons moving in the direction $\mu$
at point ${\bf x}$ and time $\eta$. The brightness function
satisfies the Boltzmann equation, which 
may be solved by expanding $\Theta({\bf x},\mu, \eta)$
in Fourier modes in space and Legendre polynomials in angle. 
The resulting components $\Theta_l(k,\eta)$ satisfy
an infinite hierarchy of ODE's, usually truncated and solved
on a computer by a so-called Boltzmann code.
The angular power spectrum $C_l$ can be obtained from the  
$\Theta_l(k,\eta)$ today ($\eta=\eta_0$) by using
\begin{equation}\label{cls}
C_l=\int_0^{\infty} dk k^2
{\langle|\Theta_l(k,\eta_0)|^2\rangle}\;.
\end{equation}
where the brackets denote ensemble averages.
In the analytical framework developed in  \cite{HS}, rather than
making use of a Boltzmann code, evolution is split into three 
regimes for which analytical solutions are found:  
tight-coupling, recombination, and free-streaming.
The idea is that before recombination photons are sufficiently
tightly coupled to behave like a perfect fluid. For a perfect
fluid the infinite series  $\Theta_l(k,\eta)$ reduces to two non-vanishing
components: the monopole $\Theta_0$ (related to the
fluid energy density), and the dipole (related to its velocity).
These satisfy perturbed perfect fluid equations, which can be 
analytically solved using the  WKB approximation scheme.
After recombination the photons free stream towards us.
Although the infinite series $\Theta_l(k,\eta)$ is now required
to describe the photons an exact solution for its evolution can be written.
In between the two epochs there is recombination. If recombination
were instantaneous one could simply glue together
the two solutions. However this is never the case. The effects
of finite recombination time can nevertheless be included in the form of 
a damping factor $D(k)$, describing Silk damping, an algorithm
for which is supplied in \cite{HS}.  A new paper by Battye has
recently suggested some potentially important corrections to this
method \cite{battye}.  The corrections could raise the overall amplitude at
small scales, but are unlikely to affect the presence or absence of the
Doppler peaks, which we address here.

\subsection{Free streaming}
The free-streaming solution is approximately
\begin{eqnarray}
\Theta_l(k,\eta_0)&\approx& [\Theta_0+\Psi](k,\eta_*)
j_l(k\Delta\eta_*) +\Theta_1(k,\eta_*){lj_{l-1}(k\Delta\eta_*)-(l+1)
j_{l+1}(k\Delta\eta_*)\over 2l+1}\nonumber\\
&&+{\int^{\eta_0}_{\eta_*}}[{\dot\Psi}-{\dot \Phi}]
j_l(k\Delta\eta)d\eta\label{freest}
\end{eqnarray}
where $\eta_*$ is the conformal time at last scattering,
$\Delta\eta=\eta_0-\eta$, $\Delta\eta_*=\eta_0-\eta_*$,
and $\Phi$ and $\Psi$ are the scalar gauge-invariant potentials
(essentially the Newtonian potential on subhorizon scales).
The first two terms represent the free streaming of spatial perturbations 
(of scale $k$) at last scattering into angular 
temperature fluctuations (of scale $l$) here and now.
The first term is the free streaming projection of spatial fluctuations
in the photon energy density (identified as $\Theta_0 +\Psi$),
and the second term the projection of fluctuations in the photons'
velocity (Doppler effect). The projectors are Bessel functions
and peak at $l\approx k\Delta\eta_*$, but with some spread.
Hence oscillations in the power spectra 
of $\Theta_0+\Psi$ and $\Theta_1$
at last scattering are
translated into oscillation in the $C_l$ spectrum, but these 
always appear smoothed in the $C_l$. It is pointed out in \cite{HS}
that the projector of the monopole term is more peaked than the projector
of the dipole. Hence oscillations in the power spectrum of the monopole
$\Theta_0$ at last scattering will be less smoothed out 
in the $C_l$ spectrum than oscillations
in the dipole $\Theta_1$. In addition to the first two terms 
there is the Integrated Sachs Wolf (ISW) term,
accounting for perturbations induced on photons in free flight
after last scattering by  non-conservative gravitational potentials.

\subsection{Tight coupling}
The equations ruling the photon fluid during tight coupling are:
\begin{eqnarray}
{\dot\Theta_0}&=&-{k\over 3}\Theta_1-{\dot \Phi}\nonumber\\
{\dot\Theta_1}&=&-{\dot R \over 1+R}\Theta_1+{k\over 1+R}\Theta_0
+k\Psi\; ,
\label{tighteqns}
\end{eqnarray}
where $R={3\over 4}{\rho_b\over\rho_{\gamma}}$ is the
scale factor normalized to $3/4$ at photon-baryon equality.
In this regime the baryons' density contrast $\Delta_b$ and velocity
$V_b$ satisfy the conditions
\begin{eqnarray}
{\dot \Delta_b}&=&{3\over 4}{\dot \Delta _{\gamma}}\label{adiab}\\
V_b&=&V_{\gamma}\; ,
\end{eqnarray}
where the photons' density contrast $\Delta_{\gamma}$ and velocity
$V_{\gamma}$ are to be found from
\begin{eqnarray}
V_{\gamma}&=&\Theta_1\\
\Delta_{\gamma}&=&4{\left(\Theta_0+h{\Theta_1\over k}\right)}\; .
\end{eqnarray}
We shall assume here that the main dynamical components are the photons
and a defect component (and to a minor extent the baryons). 
Therefore we set the CDM density contrast $\Delta_c$ to zero.
We also assume that Eqn.(\ref{adiab}) can be integrated into
$\Delta_b=(3/ 4)\Delta _{\gamma}$, that is, 
there are no entropy perturbations. These assumptions are not
a requirement for the work to be presented, but are merely
a simplification.

It may now happen that the potentials driving the fluid are 
external, or at any rate, known a priori. If this is the case then
a WKB solution
may be found for the monopole and dipole terms. For large
$k$ (say $k>0.08h^3$) this is 
\begin{eqnarray}
[{\hat\Theta}_0+\Psi](\eta_*)&=&[\Psi-\Phi](\eta_*)+
{1\over(1+R_*)^{1/4}} \bigg(
[\cos{kr_s(\eta_*)}+J(0)\sin{kr_s(\eta_*)}][{\Theta}_0+\Psi](0) +I(\eta_*)
\bigg)\label{monop}\\
{\hat\Theta}_1(\eta_*)&=&
{{\sqrt 3}\over (1+R_*)^{3/4}}\Bigg(
[1+J(\eta_*)J(0)][{\Theta}_0+\Psi](0)\sin{kr_s(\eta_*)}\nonumber\\
&+&[J(\eta_*)-J(0)][{\Theta}_0+\Psi](0)\cos{kr_s(\eta_*)}\nonumber\\
&+&J(\eta_*)I(\eta_*)-
{k\over{\sqrt 3}}{\int_0^{\eta_*}}d\eta\Phi(\eta)G(\eta)
\cos[kr_s(\eta_*)-kr_s(\eta)]\Bigg)\label{dip}
\end{eqnarray}
with 
\begin{eqnarray}
I(k,\eta)&=&{k\over{\sqrt 3}}{\int_0^{\eta}}d\eta'\Phi(\eta')G(\eta')
\sin[kr_s(\eta)-kr_s(\eta')]\\
G(k,\eta)&=&(1+R)^{-1/4}{\left(1-(1+R){\Psi\over \Phi}
+{3{\ddot R}\over 4k^2}-J^2\right)}\\
J(k,\eta)&=&{{\sqrt{3}}\over 4k}{{\dot R}\over {\sqrt{1+R}}}
\end{eqnarray}
and where $r_s(\eta)$ is the sound horizon
\begin{equation}
r_s(\eta)={1\over {\sqrt{3}}}\int^{\eta}_0 d\eta ' {1\over 
{\sqrt{1+R}}}\; .
\end{equation}
For small $k$ (say $k<0.08h^3$) one may simply set
$R=0$ in the solutions given above.

The gauge-invariant scalar potentials driving this system 
may be obtained from Einstein's equations \cite{KS}
\begin{eqnarray}\label{poteq}
k^2\Phi&=&4\pi{\left(a^2\rho\Delta_T +\rho^s+3
h v^s\right)}\label{poteq1}\\
\Phi+\Psi&=&-8\pi{\left(a^2{p\Pi\over k^2}+\Pi^s
\right)}\label{poteq2}
\end{eqnarray}
where $a$ is the scale factor, $h=\dot{a}/a$, $\rho$ ($\Delta_T$) is
the total matter density (density contrast), and $p\Pi/k^2$ is simply 
related to the nearly vanishing quadrupole of the 
photon and neutrino fluctuations. In the tight coupling regime
one may set $\Pi=0$ in a first approximation, and the total
density contrast may be written as
\begin{equation}
\rho\Delta_T=\rho_b\Delta_b+\rho_{\gamma}\Delta_{\gamma}
=\rho_{\gamma}\Delta_{\gamma}(1+R)
\end{equation}
from Equation~\ref{adiab}.
We have included
a defect component with  stress-energy tensor $\Theta_{\mu\nu}$ given by
\begin{eqnarray}
\Theta_{00}&=&\rho^s\nonumber \\
\Theta_{0i}&=&k_iv^s\nonumber\\
\Theta_{ij}&=&p^s\delta_{ij}+(k_i
k_j-{1\over 3}\delta_{ij}k^2)\Pi^s
\end{eqnarray}
as in \cite{mypaper} (but rewritten in terms
of $\Theta_{\mu\nu}$ rather than $\Theta^{\mu}_{\nu}$). 
Equations (\ref{poteq1}) and (\ref{poteq2}) have the advantage of being
elliptical, thus allowing a simple connection between sources and
potentials. They form a complete set as long as the sources are conserved.
For the defects this means:
\begin{eqnarray}
{\dot\rho}^s+h(3p^s+\rho^s)+k^2v^s&=&0\label{cons1}\\
{\dot v}^s+2hv^s-p^s+{2\over 3}k^2\Pi^s&=&0\label{cons2}\; .
\end{eqnarray}

\section{Topological defects contrasted with inflation}\label{definfl}
We now focus on the basic assumptions of
inflationary and defect theories and  isolate  the 
contrasting properties. This will allow us to write down
a defect tailored Hu and Sugiyama formalism. We shall
use the resulting formalism both in general arguments
concerning defect scenarios, and in 
rigorous calculations for particular defects.
We define the concepts of active and passive perturbations,
and of coherent and incoherent perturbations. In terms of 
these concepts inflationary perturbations are 
passive coherent perturbations.
Defect perturbations are active perturbations more or less
incoherent depending on the defect.

\subsection{Active and passive perturbations, and their different
perceptions of causality and scaling}
The way in which inflationary and defect perturbations come about
is radically different. Inflationary fluctuations 
were produced at a remote epoch, and were 
driven far outside the Hubble radius by  inflation.  The
evolution of these fluctuations is linear (until 
gravitational collapse becomes non-linear at late times), and we call
these fluctuations ``passive''.  Also, because
all scales observed today have been in causal contact since the onset
of inflation, causality does not strongly constrain the fluctuations
which result. In contrast, defect fluctuations are continuously seeded by
defect evolution, which is a non-linear process.
We therefore say these are ``active'' perturbations.  Also, the
constraints imposed by causality on defect formation  and evolution 
are much greater than than those placed on inflationary perturbations.

\subsubsection{Active and passive scaling}
The notion of scale invariance has different implications
in these two types of theory. For instance, a scale invariant gauge-invariant
potential $\Phi$ with dimensions $L^{3/2}$ has a power spectrum 
$$P(\Phi)=\langle |\Phi_{\bf k}|^2\rangle\propto k^{-3}$$ 
in passive theories (the Harrison-Zeldovich spectrum). 
This results from the fact that
the only variable available is $k$, and so the only spectrum one
can write down which has the right dimensions and does not have a 
scale is the Harrison-Zeldovich spectrum. 
The situation is different
for active theories, since time is now a variable.
The  most general counterpart to the Harrison-Zeldovich spectrum is  
\begin{equation}\label{scale}
P(\Phi) =  \eta^3F_{\Phi}(k\eta)
\end{equation}
where $F_{\Phi}$
is, to begin with, an arbitrary function of $x=k\eta$. All other
variables may be written as a product of a power of $\eta$, ensuring
the right dimensions, and an arbitrary function of $x$. 
Inspecting all equations it can be checked that it is possible to do
this consistently for all variables. All equations respect scaling
in the active sense.

\subsubsection{Causality constraints on active perturbations}\label{causal}
Moreover, active perturbations are constrained by causality, in the form of 
integral constraints \cite{trasch12,james}. These consist of 
energy and momentum conservation laws for fluctuations
in an expanding Universe. The integral constraints can be used to 
find the low $k$ behaviour of the perturbations' power spectrum, 
assuming their causal generation and evolution
\cite{traschk4}. Typically it is found that the causal creation
and evolution of defects requires that their energy $\rho^s$ and scalar
velocity $v^s$
be white noise at low $k$, but that the total energy fluctuations'
power spectrum is required to go like $k^4$. To reconcile these two facts
one is forced to consider the compensation. 
This is an underdensity in the matter-radiation  energy density
with a white noise low $k$ tail, correlated with the defect network
so as to cancel the defects' white-noise tail.

We first derive directly the low $k$ behaviour of all variables
in our problem. For low $k$ one may set $R=0$, 
$\Delta_T\approx\Delta_{\gamma}\approx 4\Theta_0$, and solve equations
(\ref{tighteqns}) coupled to (\ref{poteq1}) and (\ref{poteq2}) to find
\begin{eqnarray}
\Theta_0&=&-4\pi{\rho^s+3hv^s\over k^2+6h^2\Omega_{\gamma}}\label{th0k0}\\
a^2\rho\Delta_T&=&-6h^2{\rho^s+3hv^s\over k^2+6h^2\Omega_{\gamma}}
\label{arDk0}\\
{\cal U}=a^2\rho\Delta_T+\rho^s+3hv^s&=&{k^2(\rho^s+3hv^s)
\over k^2+6h^2\Omega_{\gamma}}\label{Uk0}\\
\Phi=-\Psi&=&4\pi{\rho^s+3hv^s\over k^2+6h^2\Omega_{\gamma}}\label{potk0}
\end {eqnarray}
\nobreak
where $\Omega_{\gamma}=\rho_{\gamma}/\rho$, and we have used $3h^2=8\pi\rho$.
This solution is in fact valid for all values of $k$ for which
the dipole $\Theta_1$ is negligible.
The lack of superhorizon correlations in the defect network requires
the power spectrum $P(\rho^s)$ to have a white noise low $k$ tail.
Energy conservation Eqn.~(\ref{cons1}) requires that $v^s$ also
have a white noise low $k$ tail.
Equations (\ref{th0k0}) and (\ref{arDk0}) show the emergence 
of the compensation: a necessary low $k$ white-noise tail 
in the radiation power spectrum exactly anticorrelated with the
defects' tail. The quantity to be cancelled is
$\rho^s+3hv^s$, and not just $\rho^s$. The density forced
to have a $k^4$ power spectrum is  ${\cal U}$,
as shown by (\ref{Uk0}), and not just a combination of
$\Delta_T$ and $\rho^s$. This is precisely the source term 
for the potential $\Phi$ which must therefore be white-noise
on large scales.  Since isotropy requires $\Pi^s$ and $p\Pi/k^2$ 
to be  constant for
small $k$, the Einstein equations 
imply that scaling active perturbations
produce scaling gauge-invariant potentials, which must be
white-noise on large scales. In particular $P(\Psi-\Phi)
=F(k\eta)\eta^3$, with $F(0)$ a non-zero constant. 
For most realistic defects $x^4F(x)$ will then have a single peak,
located 
at a value of $x\equiv x_c$ corresponding roughly to the
``coherence scale'' of the defect in question.
We will see that the place and thickness of the
peak in $x^4F(x)$ are deciding features for
the Doppler peaks induced by active perturbations.

These results can be understood by means of integral constraints.
Although gauge-invariant in their initial formulation,
integral constraints have been applied to defects in terms of 
the gauge dependent energy-momentum pseudo-tensor of \cite{steb,pen}. 
Here we go back to their original formulation, and
rewrite the defect integral  constraints in terms
of gauge-invariant variables. If a given perturbation has a compact
support then according to \cite{trasch12} the following quantity
must be conserved
\begin{equation}\label{energy}
{\cal E}=\int dV(\delta T^0_0-x^ih\delta T^0_i)\; .
\end{equation}
It can be checked that the density 
${\cal U}=(\delta T^0_0-x^ih\delta T^0_i)$ is gauge invariant
under transformations which go to zero outside the perturbation
boundary. Since the Universe
is unperturbed outside this boundary, setting all 
$local$ perturbation variables to zero in this region is a
natural  gauge restriction.  
Assuming a purely scalar perturbation, and 
integrating the second term in (\ref{energy}) by parts 
the density ${\cal U}$ can then be written as
\begin{equation}
{\cal U}=a^2 \rho\Delta_T +\rho^s+3hv^s
\end{equation}
which is precisely the source term in (\ref{poteq1}) and the 
quantity given in (\ref{Uk0}). 
Using similar arguments it can be shown that the other three
integral constraints concern integrals of the form $\int dV\,x^i
{\cal U}$.
Using these two constraints it was 
shown in \cite{traschk4} that if the perturbation field
is the sum of individual contributions with a compact support and
which are uncorrelated beyond a certain length, then the
power spectrum of the total ${\cal U}$ field has to be bound
by $k^4$ as $k\rightarrow 0$. This is precisely what we have found
by studying directly the low $k$ behaviour of the tight-coupled
equations.

The compensation can be  implemented in the form of
a compensation  factor \cite{AS}. In our language this means postulating
an equation of state for $\Delta_T$ of the form
\begin{equation}\label{fudge1}
{\cal U}=a^2 \rho\Delta_T +\rho^s+3hv^s=\gamma_c(\rho^s+3hv^s)
\end{equation}
Usually one writes
\begin{equation}\label{fudge2}
\gamma_c={1\over 1+{\left({\chi_c\over x}\right)}^2}
\end{equation}
where $\chi_c$ is the compensation scale. Equation (\ref{Uk0})
{\it suggests} that if $h\eta=\alpha$ then $\chi_c={\sqrt {6\Omega_{\gamma}}}
\alpha$. Then $\chi_c\approx 2.45$ in the radiation era. 
In work in preparation \cite{causal} we show that
$\gamma_c$ is more than just a ``fudge'' factor.
The radiation backreaction is indeed dominated by a factor
of the form of $\gamma_c$. However, secondary radiation backreaction
effects occur. A dipole driven 
monopole wave, with $k^4$ large scale behaviour and so 
not required by causality, is of particular relevance. 
It always acts so as to push the compensation scale further inside
the horizon. Hence we consider a range of $\chi_c$ going over its
limiting value of 2.45.

\subsection{Coherent, incoherent, and totally incoherent perturbations}
Active perturbations may also differ from
inflation in the way ``chance'' comes into the theory. 
Randomness occurs in inflation only when the initial 
conditions are set up. Time evolution is linear and
deterministic, and may be found by
evolving all variables from an
initial value equal to the square root of 
their initial variances. By squaring the
result one obtains the variables' variances at any time.
Formally this results from unequal time correlators of the form
\begin{equation}\label{2cori}
{\langle\Phi({\bf k},\eta)\Phi({\bf k'},\eta ')\rangle}=
\delta({\bf k}-{\bf k'})\sigma({\Phi}(k,\eta))\sigma
({\Phi}(k,\eta')),
\end{equation}
where $\sigma$ denotes the square root of the power spectrum $P$.
In defect models however, randomness may intervene in the time
evolution as well  as the initial conditions. 
Although deterministic in principle, 
the defect network evolves as a result of a 
complicated non-linear process.
If there is strong non-linearity, a given mode will be ``driven'' 
by interactions with the other modes in a way which will force
all different-time correlators to zero on a time scale
characterized by the ``coherence time'' $\tau_c(k,\eta)$.
Physically this means that one has to perform a new ``random'' draw 
after each coherence time in order to
construct a defect history \cite{us}. 
The counterpart to (\ref{2cori}) for incoherent perturbations is
\begin{equation}\label{pr0}
{\langle\Phi({\bf k},\eta)\Phi({\bf k'},\eta ')\rangle}=
\delta({\bf k}-{\bf k'}) P({\Phi}(k,\eta),\eta'-\eta)\; .
\end{equation}
For $|\eta'-\eta| \equiv |\Delta\eta|> \tau_c(k,\eta)$
we have $P({\Phi}(k,\eta),\Delta\eta)=0$. For $\Delta\eta=0$,
we recover the power spectrum $P({\Phi}(k,\eta),0)=P({\Phi}(k,\eta))$.
If there is active scaling we have seen that $P({\Phi}(k,\eta))=
\eta^3F_{\Phi}(x)$. Also the correlation
$$
{\rm cor} (\Phi(k,\eta),\Phi(k,\eta'))={
{\langle\Phi(k,\eta)\Phi(k,\eta ')\rangle}\over
\sigma (\Phi(k,\eta))\sigma (\Phi(k,\eta'))}
$$
must be a function of only $x=k\eta$ and $x'=k\eta'$. Let's
call it ${\cal C}_{\Phi}(x,x')$. Since there is no
time translational invariance, this 
function is not necessarily dependent only
on $x-x'$. Hence, if there is active scaling, we can write
\begin{equation}\label{twop}
{\langle\Phi({\bf k},\eta)\Phi({\bf k'},\eta ')\rangle}=
\delta({\bf k}-{\bf k'}) (\eta^3 F_{\Phi}(x))^{1/2}
(\eta'^{3} F_{\Phi}(x'))^{1/2} {\cal C}_{\Phi}(x,x').
\end{equation}
Detailed understanding of the effects of incoherence on the structure
of Doppler peaks requires knowledge of the coherence function
${\cal C}(x,x')$. However two approximation schemes have been implemented.
In one effective coherence was assumed so that  ${\cal C}(x,x')=1$.
This was done in calculations for texture  models \cite{neil,durrer},
where coherent statistics were assumed. In another \cite{us}
it was assumed that 
\begin{equation}\label{pr}
{\langle\Phi({\bf k},\eta)\Phi({\bf k'},\eta ')\rangle}=
\delta({\bf k}-{\bf k'}) \delta (\eta-\eta')
P_r({\Phi}(k,\eta)),
\end{equation}
in which
\begin{equation}\label{prpk}
P_r({\Phi}(k,\eta))={{\int}}\,d\Delta\eta
P({\Phi}(k,\eta),\Delta\eta)
\end{equation}
is the time-integrated power spectrum \cite{AS}.   
This is far less naive than just setting ${\cal C}(x,x')=\delta (x-x')$
in (\ref{twop}) and draws on the methods of \cite{AS}. 
It should be a good approximation whenever convolving
$P({\Phi}(k,\eta),\Delta\eta)$ with functions
which vary slowly at the scale of $\tau_c(k,\eta)$.
Clearly both approximations have drawbacks, and it is up to 
detailed calculations to decide which
of the two approximations is better, if any, in each particular situation.

We shall label as coherent, incoherent,  and totally incoherent 
the perturbations satisfying
(\ref{2cori}), (\ref{pr0}),  and (\ref{pr})  respectively. This feature
changes the way the average $C^l$ are computed, 
resulting in  a striking qualitative difference 
in the structure of Doppler peaks.  

A comment should be added concerning the possibility of different
coherence properties among the different components which
act as sources for
the potentials $\Phi$ and $\Psi$ (cf. Eqns. (\ref{poteq1})
and (\ref{poteq2})). Indeed it could happen that say,
radiation behaved coherently while defects behaved incoherently.
In such a case the potentials coherence properties would naturally
mix their sources coherence properties in a fashion easily predictable
from Einstein's equations. However, as we
explain in Section~\ref{looph}, it is a good approximation 
to neglect the $a^2\rho\Delta_T$ contribution to the potentials
except for the compensation. The compensation, on the other hand,
is given by the large scale solution found in Section~\ref{causal}.
It should be noted that Eqn.(\ref{arDk0}) applies for each realization,
implying perfect correlation between defects and compensation.
It follows that the compensation
has the same coherence properties as the defect network. Therefore
within the  approximation spelled out further in Section~\ref{looph}, 
the potentials have the same coherence properties as the defects.

\section{A defect-tailored Hu and Sugiyama formalism}\label{hsdefect}
In scenarios in which perturbations are driven by a topological defect
network, a number of simplifications and extensions apply to the Hu
and Sugiyama formalism. Making use of active scaling and the low $k$
constraints (\ref{th0k0}) to (\ref{potk0}) one may prove that one may
drop all but the convolution terms in (\ref{monop}) and (\ref{dip}).
In fact they will be suppressed relative to the convolution terms by a 
factor of order $(\eta_{ph}/\eta_*)^{3/2}$, where $\eta_{ph}$ is the
conformal time of the phase transition. Then the multipoles 
$\Theta_l(k,\eta_0)$ may be written in the simplified form
\begin{eqnarray}
\Theta_l(k,\eta_0)&=&{\int_0^{\eta_*}}d\eta\, k\Phi(k,\eta)
G(k,\eta){\left( D^l(k,\eta)+V^l(k,\eta)\right)}+\nonumber\\
&+& (\Psi-\Phi)(\eta_*)j_l(k\Delta\eta_*)+
{\int_{\eta_*}^{\eta_0}}({\dot\Psi}-{\dot\Phi})j_l(k\Delta\eta)
\end{eqnarray}
with  
\begin{eqnarray}
D^l(k,\eta)&=&{j_l(k\Delta\eta_*)\over{\sqrt{3}}(1+R_*)^{1/4}}
\sin(kr_s(\eta_*)-kr_s(\eta))\nonumber \\
V^l(k,\eta)&=&{lj_{l-1}(k\Delta\eta_*)-(l+1)j_{l+1}(k\Delta\eta_*)
\over (2l+1) (1+R_*)^{3/4}}\times\nonumber \\
&\times&[ J(\eta_*)\sin(kr_s(\eta_*)-kr_s(\eta))
-\cos(kr_s(\eta_*)-kr_s(\eta))]\; .
\end{eqnarray}
If one assumes coherence then one may simply compute the 
$ \Theta_l(k,\eta_0)$ with this formula, replacing every variable by
the square root of its power spectrum. By squaring the result one then
obtains ${\langle |\Theta_l(k,\eta_0)|^2\rangle}$, providing the
$C_l$ spectrum by means of (\ref{cls}). If the perturbation is 
incoherent then the H+S formalism must be modified
at this step. For totally incoherent perturbations (imposing (\ref{pr})) one 
has instead: 
\begin{eqnarray}\label{clicr}
C_l&=&\int dk{\int_0^{\eta_*}} d\eta k^4P_r(\Phi(k,\eta)G(k,\eta))
{\left({D^l(k,\eta)+V^l(k,\eta)}\right)}^2+\nonumber\\
&+&\int dk k^2 P(\Psi-\Phi)(\eta^*)j_l^2(k\Delta\eta_*)
+\int dk k^3{[V^lj_l(k\Delta\eta_*)\sigma^r(\Phi G)\sigma^r(\Psi
-\Phi)](\eta^*)}
\nonumber \\
&+&\int dk{\int_{\eta_*}^{\eta_0}} 
d\eta k^2 P_r({\dot \Psi}-{\dot\Phi})j^2_l(k\Delta\eta)\; .
\end{eqnarray}
For incoherent perturbations the problem is always more complicated
numerically, as one has to perform two integrals in time and
one in $k$.

Even before performing any calculations one may expect striking
differences between defects and inflationary Doppler peaks.
For active perturbations the convolution terms dominate over
the primordial terms in the monopole and dipole terms. 
Since the phase $kr_s(\eta_*)$ enters differently
in the primordial and convolution terms one may expect 
differences in the Doppler peaks' position.
Also, in obtaining the $C_l$ for coherent active perturbations
one integrates an oscillatory function 
and then squares the result. Some oscillatory structure can always
be expected in the $C_l$. For 
totally incoherent active perturbations,
on the contrary,  one simply integrates
the square of an oscillatory function. This contains a DC level
as well as an AC level. These will compete,
leaving it up to other details of the incoherent perturbation
to decide whether or not there are secondary oscillations.
We will expand in great detail on these two points.

\subsection{Possible loopholes}\label{looph}
Eqns. (\ref{cls}), (\ref{freest}), 
(\ref{monop}), (\ref{dip}), (\ref{poteq1}), and  (\ref{poteq2})
seem to imply that topological defects' $C_l$ spectra may be
analytically computed just from knowledge of
the defect two-point functions
\begin{equation}
  \label{data}
  {\langle \Theta_{\mu\nu}(\eta) \Theta_{\alpha\beta}(\eta')\rangle}.
\end{equation}
This includes knowledge of time-time correlators, and also cross-correlators
involving different stress-energy components. However there are two
loopholes in this statement which we now spell out.

In the H+S formalism one assumes that the potentials are external to
the photon-baryon fluid, or at least, are known a priori.  
The defect potentials are indeed external, since defects evolve according
to their own independent dynamics (the so-called stiff defect assumption
\cite{steb}). 
However we know that we cannot neglect $\Delta_{\gamma}$ on large scales,
as the compensation is of the same order of magnitude as the defect
perturbations themselves. Since $\Delta_{\gamma}$ acts as a source for
its own driving potentials we have a loophole in the formalism,
as we cannot account for radiation backreaction effects.
This is not necessarily a problem, as we have an explicit
low $k$ solution for the radiation (Eqns. (\ref{th0k0}) to 
(\ref{potk0})). Hence we may account for the nonignorable
radiation backreaction in the form of a  factor $\gamma_c$,
as in Eqn.(\ref{fudge2}).
Thus we include radiation backreaction 
where it is required by causality,
and ignore it otherwise. Many of the uncertainties
in the backreaction may be mimicked by leaving $\chi_c$ as a free
parameter. We shall see that the  particular Doppler peak features we are
interested in are insensitive to this uncertainty.
In future work \cite{causal} we 
will write down solutions for the tight-coupling equations driven by a defect
source, which incorporate exact radiation backreaction.
These show that  $\gamma_c$ is in fact the leading order
backreaction effect. In another approach \cite{realizations}
we will simply solve the differential equations (\ref{tighteqns})
directly. Although this approach can cope with the compensation exactly,
it has the disadvantage of requiring  a Monte-Carlo technique.

A second loophole concerns the measurement of the two-point
correlators (\ref{data}) from simulations. It may happen
that we can only measure one term in the overall defect associated
stress-energy. For instance
a full account of cosmic string evolution requires consideration
of evolution of tiny loops and
the gravitational radiation they dissolve into. These
are usually overlooked in string simulations. 
However it is  extremely
important to keep track of every defect stress-energy term,
as one needs them all to ensure energy conservation, without which
Eintein's equations are not integrable. If however one assumes 
that the overlooked terms 
are a minor contribution to the total stress-energy 
tensor, then they can be neglected in our formalism. Equations
(\ref{poteq1}) and (\ref{poteq2}) can be solved separately for every
stress-energy component. If gravitational radiation is negligible
then so are its induced potentials $\Phi$ and $\Psi$. This statement
remains true even though one needs the gravitational radiation
component to integrate the whole set of Einstein's equations.

These loopholes concern calculations of $C_l$ spectra for concrete
motivated defect scenarios (such as cosmic strings or textures).
They do not affect the general analysis performed in Section~\ref{genqual}.

\section{Doppler peaks for a generic defect}
\label{genqual}
We now undertake a  qualitative  general analysis of peak position and
secondary oscillation strength in the monopole power spectrum
at last scattering.
We input qualitative perturbation features, as general as possible, 
for which we output only qualitative peak  features.
The idea is to find out which novelties in defect
peak features are generic defect properties as opposed to
mere accidents pertaining to particular defect models. 
Also we may learn when and why the coherent and totally incoherent 
approximations are too crude to be useful.
\begin{figure}
\begin{center}
    \leavevmode
        {\hbox %

{\epsfxsize = 8cm\epsfysize=8cm
    \epsffile {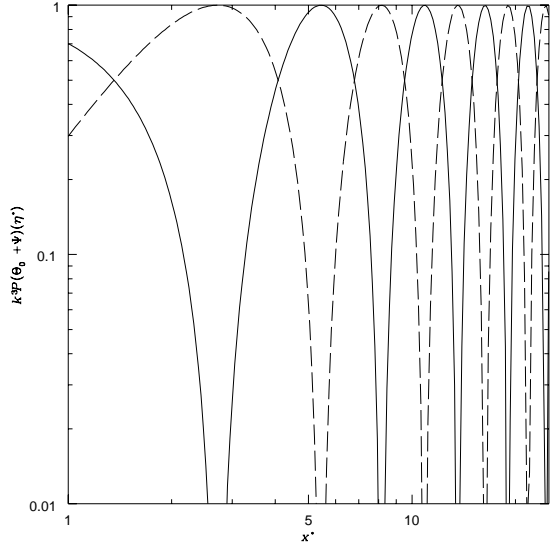} }}
\end{center}
\caption{The dimensionless power spectrum of the monopole term
$k^3 P(\Theta_0+\Psi)(\eta_*)$ plotted as a function of 
$k^*$  for adiabatic (line) and isocurvature
(dashed) passive fluctuations.}
\label{infl}
\end{figure}

The free-streaming solution (\ref{freest}) suggests interpreting
$(\Theta_0+\Psi)(\eta_*)$ as the intrinsic anisotropy $\delta_{\gamma}$,
that is, the anisotropy due to photon energy density fluctuations
at last scattering.
This anisotropy has the sharpest free-streaming conversion from $k$ to $l$
and is responsible for the overall Doppler peak features.
In particular the Doppler peak's position and the structure of secondary
oscillations may be inferred from the power spectrum
$P(\delta_{\gamma})(\eta_*)$. However
it should be borne  in mind that oscillations
in $P(\delta_{\gamma})$ overestimate oscillations in the
$C_l$. If we stick to low values of $\Omega_b h^2$ then the position
and structure of the peaks is only mildly dependent on these parameters.
In this Section we neglect this dependence altogether by implementing an
approximation where $R=0$, $r_s(\eta)=\eta/ {\sqrt 3}$.
This is a simplification for illustrative purposes
and not a formalism requirement. Then, from (\ref{monop}), we have 
\begin{equation}
(\Theta_0+\Psi)(\eta_*)=(\Psi-\Phi)(\eta_*)+{k\over {\sqrt 3}}
\int_0^{\eta_*}d\eta\,(\Phi-\Psi)\sin {{k(\eta_*-\eta)\over {\sqrt 3}}}
\end{equation}
If there is active scaling, in the sense of Eqn.(\ref{twop}),  one has 
\begin{equation}
{\langle(\Psi-\Phi)(\eta)(\Psi-\Phi)(\eta')\rangle}=
(\eta^3 F(x))^{1/2}(\eta'^3 F(x'))^{1/2}{\cal C}(x,x').
\end{equation}
with $x=k\eta$, $x'=k\eta'$. Here $F(x)$ is the structure function
(as in \cite{AS}) or scaling factor (as in Eqn.(\ref{scale}))
of the potential $\Psi-\Phi$,
\begin{equation}
P(\Psi-\Phi)=\eta^3F(x)
\end{equation}
and ${\cal C}$ is its coherence function.
The dimensionless power spectrum 
$k^3P(\delta_{\gamma})(\eta_*)$ is then only dependent on $x_*=k\eta_*$
and may be computed from
\begin{eqnarray}\label{pdelta}
k^3P(\Theta_0+\Psi)(\eta_*)&=&x^3_*F(x_*)
-{2\over {\sqrt 3}}(x_*^3F(x_*))^{1/2}
{\int_0^{x_*}}dx\, {\cal C}(x,x_*)(x^3F(x))^{1/2}
{\sin {x_*-x\over {\sqrt 3}}}+\nonumber\\
&+&{1\over 3}\int_0^{x_*}\int_0^{x_*}dx\,dx'\,
{\left( (x^3F(x))^{1/2}{\sin {x_*-x\over {\sqrt 3}}}\right)}
{\left( x\rightarrow x'\right)}{\cal C}(x,x')
\end{eqnarray}
If we use a coherent approximation (${\cal C}=1$) this becomes
\begin{equation}\label{pdeltac}
k^3P(\Theta_0+\Psi)(\eta_*)=
{\left((x^3_*F(x_*))^{1/2}-{1\over {\sqrt 3}}
{\int_0^{x_*}}dx\, (x^3F(x))^{1/2}
\sin{ {x_*-x\over {\sqrt 3}}}\right)}^2
\end{equation}
If we use a totally incoherent approximation one has instead
\begin{equation}\label{pdeltai}
k^3P(\Theta_0+\Psi)(\eta_*)=x^3_*F(x_*)+
{1\over 3}{\int_0^{x_*}}dx\, x^4F_r(x)
\sin^2 {{x_*-x\over {\sqrt 3}}}
\end{equation}
in which $F_r$ is the structure function for the time-integrated
power spectrum of $\Psi-\Phi$ as defined by (\ref{pr}) and 
(\ref{prpk}):
\begin{equation}
P_r(\Psi-\Phi)=\eta^4F_r(x)
\end{equation}
The rough peak structure in  passive theories may be obtained by
dropping all but the primordial terms \cite{HS}, thereby ignoring
sub-horizon processing. Thus
for Harrison-Zeldovich initial conditions one has:
\begin{eqnarray}
k^3P(\Theta_0+\Psi)(\eta_*)&\propto&{\cos^2 
{x_*\over {\sqrt 3}}}\label{adiaP}\\
k^3P(\Theta_0+\Psi)(\eta_*)&\propto&{\sin^2 
{x_*\over {\sqrt 3}}}\label{isocP}
\end{eqnarray}
for adiabatic and isocurvature passive fluctuations. 
For reference we have plotted these spectra in Figure~\ref{infl}.
The maxima are at scales $x^*_{m}/{\sqrt 3}=m\pi$ and
$x^*_{m}/{\sqrt 3}=(m+1/2)\pi$, respectively. There are true
zeros in the power spectrum, which get smoothed out into valleys
in  the $C_l$ spectrum. These
valleys are further filled by the dipole, but since this is suppressed,
the oscillatory structure survives.
The free-streaming monopole projector is simply a Bessel function
$j_l(x)$, which  has its main peak at $x\approx l$. Then 
the peaks in $k^3 P(\Theta_0+\Psi)(\eta_*)$ will be
converted into angular scales
$l_m\approx(\eta_0/\eta_*)x_{m}^*$. An improvement on this formula may be
obtained by noticing that the peak of $j_l(x)$ is in fact at
$x\approx l+0.8 l^{1/3}$.

\subsection{A representative  generic defect}\label{gendef}
We now cut a representative section through the infinite dimensional defect
parameter space. The physical
inputs into (\ref{pdelta}) are the coherence function ${\cal C}(x,x')$
and the potential structure function. Both may be obtained from the
sources via Einstein's equation:
\begin{equation}
k^2(\Phi-\Psi)=8\pi(a^2\rho\Delta_T+\rho^s+3
h v^s+k^2\Pi^s)
\end{equation}
Then if $F_{\rho^s}$ is the defect energy structure function
\begin{equation}
  P(\rho^s)=F^2_{\rho^s}/\eta
\end{equation}
one has
\begin{equation}\label{gamadef}
x^4F(x)=(8\pi)^2\gamma_c^2F^2_{\rho^s}\Gamma^2
\end{equation}
in which we use a compensation of the form (\ref{fudge1}),
and have assumed that the defect's $v^s$ and $\Pi^s$ are related 
to its energy via equations of state, leading to the factor $\Gamma$.
$F_{\rho^s}^2\Gamma^2$ must go to white noise at low $k$, and for 
realistic defects it falls-off like $1/x^n$ at large $k$. 
Then $F(x)$ must be white noise at low $x$, tailing off like $1/x^{n+4}$
at high $x$. $x^4F(x)$ 
will then have a peak at  a turn-over scale close to
$x_c \equiv 2\pi\eta/\xi_c$, where  $\xi_c$ is 
approximately the coherence length of 
the defect (the wavelength where $F_{\rho^s}^2\Gamma^2$ switches
from white noise to power law fall-off). This means that
$x^3F(x)$ and $x^4F_r(x)$ will also have a peak, typically not
very far off the peak in the potential source structure function $x^4F(x)$.
Inspecting Eqns.~\ref{pdelta}, \ref{pdeltac}, and \ref{pdeltai}
we see that more than the exact form of the potential structure functions,
it is the place and thickness of this peak that will determine
the Doppler peak structure. For the purpose of the qualitative discussion
to be carried out we then assume for definiteness that
for our generic defect 
\begin{equation}
F(x)=\exp{x^2\over 2\sigma^2}
\label{expf}
\end{equation}
placing the peak of $x^4F(x)$ at $x_c=2\sigma$. This form allows
us to consider a one parameter family of structure functions for which
both the peak position and the width are determined by the single
parameter $\sigma$.  In the familiar cases this ``one scale'' feature of the
structure function is realistic, but it is worth noting that exotic
cases could provide exceptions to the straightforward analysis which
follows from (\ref{expf}).
It is an important question
of principle to know how small $x_c$ may be before causality is violated.
The value $x_c\approx 2.7$ was suggested in \cite{james}.  
Recently, a
more systematic analysis of this question has been given by
Turok\cite{causal-ngt} who gets similar results. 

Strictly speaking the structure function (\ref{expf}) violates causality.
>From (\ref{gamadef}) it implies a structure function for $\rho^s$
which once inverted into real space reveals vanishingly small,
but not strictly zero superhorizon correlations. If one wants to
be pedantic about this, one may simply set the real space structure
function to zero outside the horizon, reinvert to Fourrier space, 
and find the corresponding modification to $F(x)$. 
Depending on how smoothly the cut-off is done a set of sub-dominant
oscillations may or may not appear in the modified causal $F(x)$
(as stated in \cite{causal-ngt}).
In any case, the correction in $F(x)$ is very small, and it certainly does
not affect the integrals leading to the $C_l$ spectra 
needed for the sake of our
argument. For this reason we shall ignore this detail in the rest of our 
discussion. 
 
For a compensation satisfying (\ref{fudge1}) 
the compensation coherence time is the same as the defect
coherence time. Hence the potential coherence function is also the defect
coherence function. We will assume that our generic defect has a coherence
function of the form
\begin{equation}
{\cal C}(x,x')=\exp{(x-x')^2\over 2\tau_c^2}
\end{equation}
in which the FWHM coherence time is given by $\theta_c
\approx 2.35 \tau_c$.

\subsection{The peak position for active perturbations}
\begin{figure}[t]
\begin{center}
    \leavevmode
        {\hbox %
{\epsfxsize = 10cm\epsfysize=10cm
    \epsffile {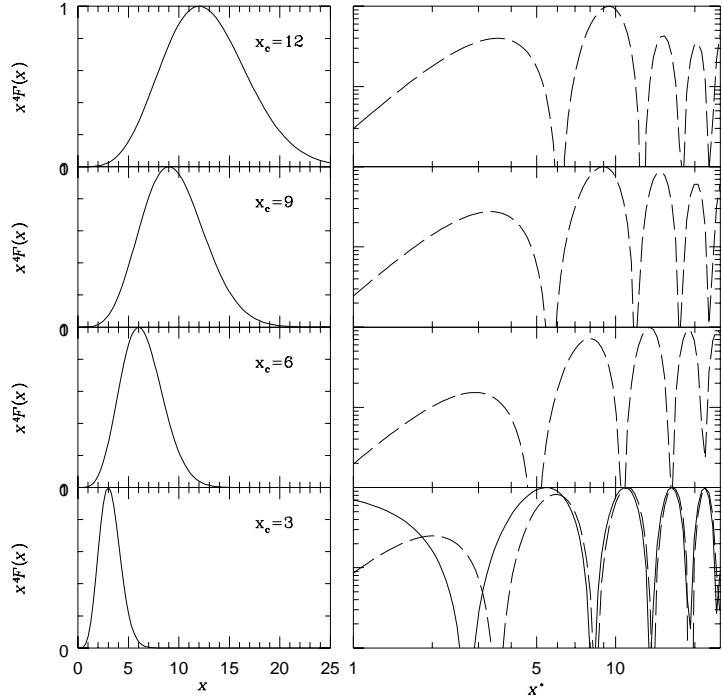} }}
\end{center}
\caption{Four potential source structure functions ($F(x)x^4$ where $F(x)$
is the potential structure function) and their corresponding
spectra $k^3 P(\Theta_0+\Psi)(\eta_*)$ (dashed lines) assuming coherence. 
The continuous line in the bottom diagram represents a passive adiabatic
spectrum.
For a sufficiently small $\xi_c$ one obtains the 
adiabatic structure of peaks. As $\xi_c$
increases one may obtain the isocurvature peaks. However one cannot
realistically push the peaks much further to the right without destroying
the symmetry of the secondary peaks.}
\label{fig2}
\end{figure}
If the peak in $x^4F(x)$ is sufficiently thin,
then from (\ref{pdelta}), (\ref{pdeltac}), and (\ref{pdeltai}), 
the monopole spectrum peaks will be at 
$x^*_m=(m-1/2)\pi{\sqrt 3}+x_c$ for coherent, incoherent, and 
totally incoherent fluctuations. Then active perturbations merely apply
a phase shift of value $x_c-\pi{\sqrt 3}/2$ to an adiabatic type of
spectrum (cf. Eqn.~\ref{adiaP}).

For $x_c\approx 2.7$ (not impossible,
but probably unrealistic because it is 
very close to the smallest
turnover point allowed by causality \cite{james,causal-ngt}) 
the monopole peaks are at the adiabatic
positions. {As $x_c$ increases from the adiabatic position} the peaks 
are shifted to smaller scales.
For $x_c\approx 5.4$ they are out of phase with the adiabatic peaks
(as in \cite{neil}). For $x_c>8.5$ the peaks start only
in the adiabatic secondary peaks region. For standard 
values of $\Omega_b$ and $h$ these three cases would place the
main ``Doppler peak'' at $l\approx 230$, $350$, and $500$, respectively. 
Therefore the placing of the peaks is {\it not} a generic
feature of active fluctuations. 
Active perturbations simply add an extra parameter on which the
Doppler peaks position is strongly dependent. In general we should expect
that for the same $\Omega$, $\Omega_b$, and $h$, active perturbations
will apply to the predicted CDM adiabatic peak position
a shift of the form
\begin{equation}
  \label{shift}
  l\rightarrow
l+{\eta_0\over \eta^*}{\left(x_c -{\pi{\sqrt3}\over 2}\right)}
\end{equation}
The secondary peaks' separation is not changed, in a first approximation.
This is to be contrasted with non-flat inflationary models 
where $C_l(\Omega=1)$ is taken into $C_{l\Omega^{-1/2}}$.
The defect shift is additive whereas the low-$\Omega$ shift 
is multiplicative, a striking difference that should always allow us to
distinguish between low $\Omega$ CDM and $\Omega=1$
high-$x_c$ defects.

If the peak in $x^4F(x)$ is thick then the issue of coherence comes
into play but all we have said still applies if the perturbation
is exactly coherent. This is illustrated in Fig.~\ref{fig2} using the
generic model defined in Section~\ref{gendef}.  
A formula for the peaks of $P(\Theta_0 +\Psi)(k,\eta_*)$ 
for more general structure functions of coherent active perturbations 
can be obtained by computing the structure function ``primitives'':
\begin{eqnarray}
F_1(x_*)&=&{\int^{x_*}_0}dx\,x^{3/2}{\sqrt {F(x)}}\cos(x/{\sqrt 3})
\nonumber \\
F_2(x_*)&=&{\int^{x_*}_0}dx\, x^{3/2}{\sqrt {F(x)}}\sin(x/{\sqrt 3})
\label{prim}
\end{eqnarray}
and noting that 
\begin{equation}
k^{3/2}(\Theta_0+\Psi)=(x^3_*F(x_*))^{1/2}
-F_1(x_*)\sin{x_*\over {\sqrt 3}}+F_2(x_*)\cos{x_*\over {\sqrt 3}}
\end{equation}
We see that defects mix the adiabatic and isocurvature modes
(cf. (\ref{adiaP}) and (\ref{isocP})). The
defect structure function primitives determine the mixing proportions,
and therefore the shift applied to the peaks.
For peaks occurring when $x^3_*F(x_*)\approx 0$ the peaks' position
is given by solutions to the equation
\begin{equation}
\tan {x_*\over {\sqrt 3}}=-{F_1(x_*)\over F_2(x_*)}\label{posi}
\end{equation}

\subsection{Secondary oscillations for totally incoherent perturbations}
\begin{figure}[t]
\begin{center}
    \leavevmode
        {\hbox %

{\epsfxsize = 10cm\epsfysize=10cm
    \epsffile {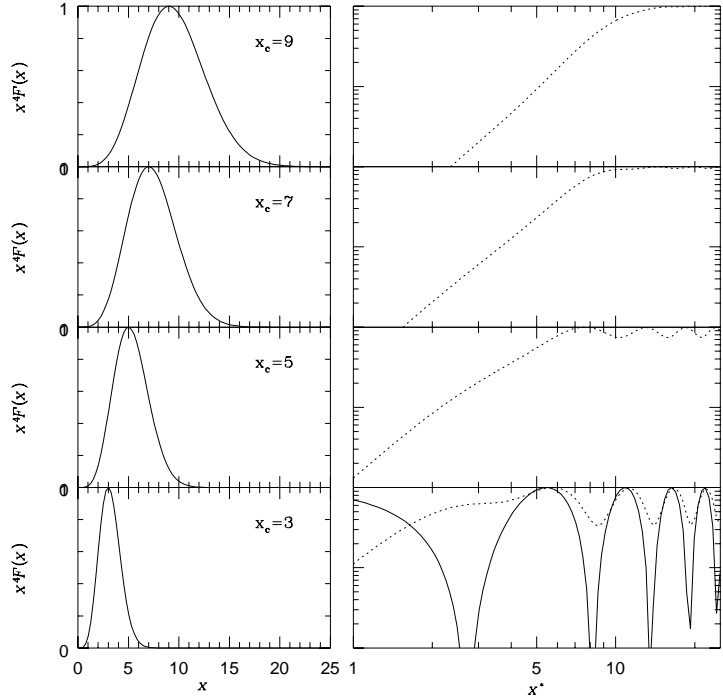} }}
\end{center}
\caption{Four potential source structure functions ($F(x)x^4$ where $F(x)$
is the potential structure function) and their corresponding
spectra $k^3 P(\Theta_0+\Psi)(\eta_*)$ (plotted with points)
assuming total incoherence. 
The continuous line in the bottom diagram represents a passive adiabatic
spectrum.
One may  obtain (softer) secondary oscillations 
at the adiabatic position for totally incoherent perturbations. 
As the spectrum shifts to the right 
(larger $x_c$) the secondary oscillations disappear
very quickly. }
\label{fig3}
\end{figure}

For totally incoherent perturbations with a thick $x^4F(x)$ peak, secondary
oscillations may never show up. All we have said on peak  position
still applies to the main peak position (it is not difficult to rewrite
(\ref{prim}) and  (\ref{posi}) for totally incoherent perturbations). 
However secondary oscillations are erased for large $x_c$.
The situation is illustrated in Fig.~\ref{fig3} for the generic 
defect defined in Sec.~\ref{gendef}, using ${\cal C}(x,x')=\delta(x-x')$
and $F_r(x)=F(x)$. Although there are 
never true zeros in $P(\Theta_0+\Psi)(x^*)$ it is 
possible to obtain significant oscillations if the main
peak is at the adiabatic position. In this case the structure function
is so narrow that the defects only source perturbations for a brief
``impluse'' time for each scale. 
However the secondary oscillations disappear very
quickly as the main peak approaches the isocurvature position,
and the structure function correspondingly broadens.
For $x_c>7$ there are no significant secondary oscillations.
The absence of secondary  oscillations is therefore not a prediction
of incoherence, as even defect's total incoherence is not sufficient
for the secondary oscillations to be erased.

\begin{figure}[t]
\begin{center}
    \leavevmode
        {\hbox %
{\epsfxsize = 10cm\epsfysize=10cm
    \epsffile {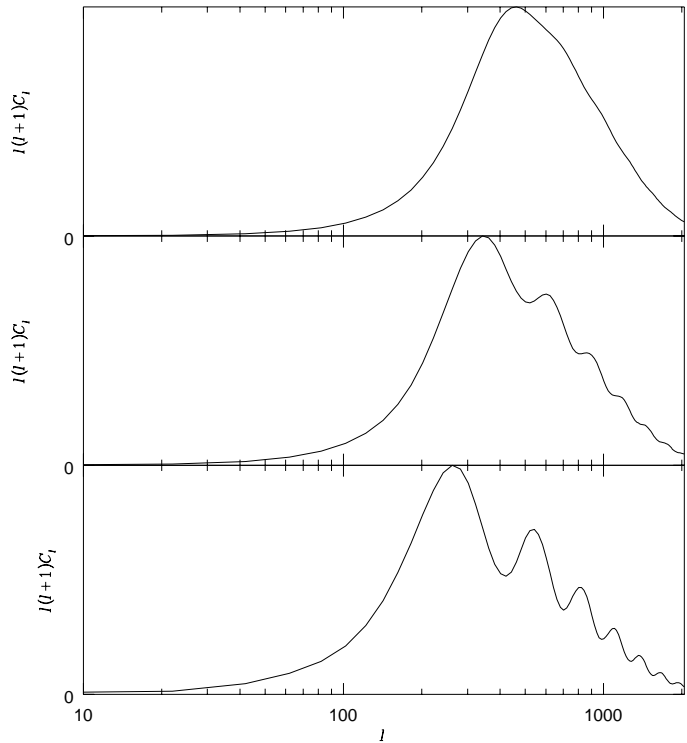} }}
\end{center}
\caption{The $C_l$ spectrum for three exponential structure functions
assuming incoherence and dropping the ISW term only. We see that 
incoherence is not by itself sufficient to erase the secondary 
oscillations. A 
totally incoherent defect with a primary near the adiabatic position
would exhibit secondary oscillations. }
\label{fig4}
\end{figure}

The low-$x_c$ incoherent oscillations present
in $P(\Theta_0+\Psi)(x^*)$ are strong enough to survive in the $C_l$.
In Fig.\ref{fig4} we show the result of solving the full H+S
algorithm for three of the totally incoherent models in Sec.~\ref{gendef}. 
We have included the monopole, dipole, and monopole-dipole (interference)
terms, but dropped the ISW term. We took $\Omega_b=0.05$,
and $h=0.5$. For the marginal value 
$x_c\approx 2.7$ one may reproduce the rough
sCDM features even with a totally incoherent defect, and with
a realistic non-zero value for $\Omega_b$. This is somewhat in
contradiction with the work of Hu and White \cite{HW},
a fact we examine more closely in work in preparation \cite{definfl}.
This example also generalizes Turoks' ``confusing defect''
\cite{causal-ngt}, as it shows that confusing defects and sCDM
is not only possible, but also that it can be done with defects with any
incoherence properties.

These results can be understood from the fact that 
if the structure function is thin, then each mode
is only active for a  short time. The defect incoherence
may not then have a chance to manifest itself in the monopole
oscillations it drives. This situation
may realistically be realized for low $x_c$ defects only.
If $x_c$ is high 
each mode is then active for several expansion times, during
which it incoherently kicks the photon plasma. The secondary oscillations
are then erased or at least softened. 

We may be more quantitative and define 
the strength of secondary oscillation as the difference between
power spectra maxima and minima in units of the maximum value:
\begin{equation}
\omega_m={(k^3P(\delta_{\gamma}))(k_m,\eta_*)-
(k^3P(\delta_{\gamma}))(k_n,\eta_*)
\over (k^3P(\delta_{\gamma}))(k_m,\eta_*)}
\end{equation}
where $k_m$ and $k_n$ are adjoining maxima and minima in the monopole 
power spectrum.
For passive or coherent active perturbations $\omega_m=1$.
Approximating the peak by a step function with width
$\delta x_c$ (not too large) an estimate of $\omega_m$ 
for totally incoherent active perturbations may be obtained.
In this approximation the dominant second term in (\ref{pdeltai})
becomes
\begin{equation}\label{pdeltaap1}
k^3P(\Theta_0+\Psi)={1\over 6}{\left(
1-{\sin{(\delta x_c/{\sqrt 3})}\over  \delta x_c/{\sqrt 3}}
\cos {2{x_*-x\over {\sqrt 3}}}\right)}
\end{equation}
and so
\begin{equation}\label{epest}
\omega\approx{2\sin(\delta x_c/{\sqrt 3})\over 
\delta x_c/{\sqrt 3}+\sin(\delta x_c/{\sqrt 3})}\; .
\end{equation}
As $\delta x_c\rightarrow 0$ one has $\omega\rightarrow 1$,
and so the secondary oscillations of totally incoherent perturbations
are indistinguishable from coherent or passive oscillations.
For, say, $\delta x_c\approx 5.2$ the oscillations are a mere undulation
with $\omega=0.1$.

\subsection{Secondary oscillations for incoherent perturbations}
\label{qualinc}
\begin{figure}[t]
\begin{center}
    \leavevmode
        {\hbox %
{\epsfxsize = 13cm\epsfysize=13cm
    \epsffile {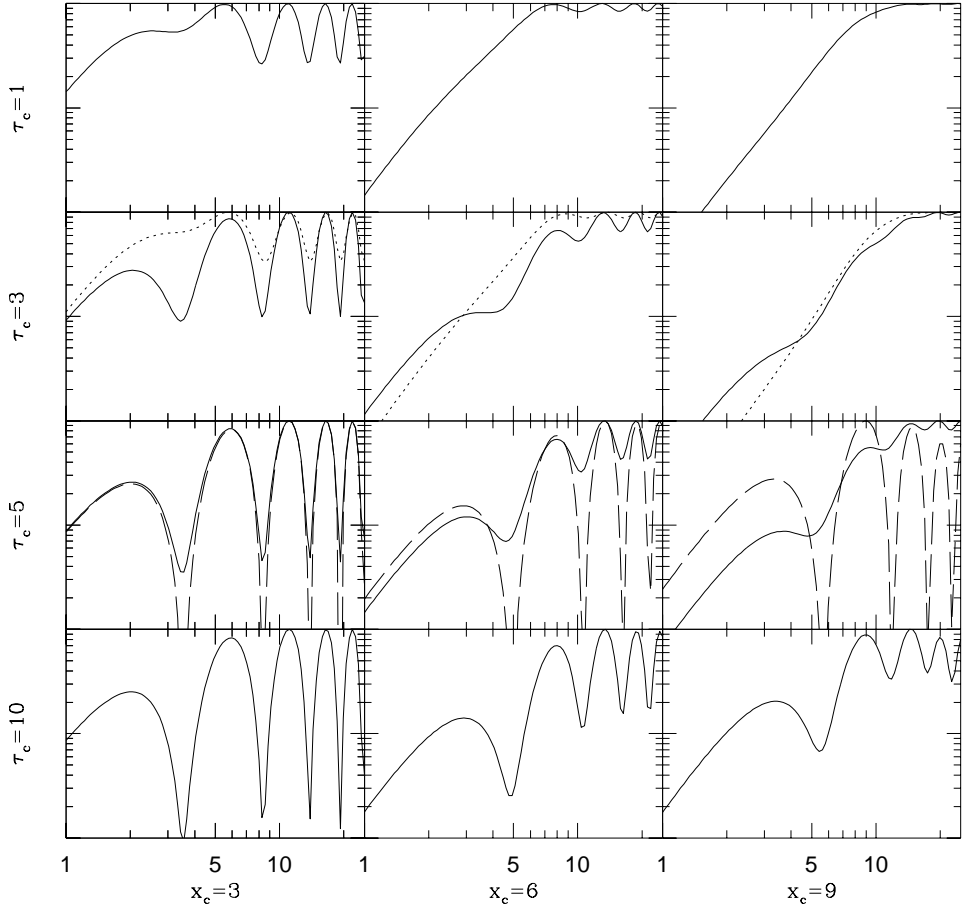} }}
\end{center}
\caption{A grid of models as in Sec. 5.1,
for various values
of $x_c$ and $\theta_c\approx 2.35\tau_c$. The dotted line on
the second row corresponds to a totally incoherent approximation,
and the dashed line in the third row to a coherent approximation. 
}
\label{fig5}
\end{figure}

If a perturbation is incoherent, with finite coherence
time $\theta_c$, then the strength of its
secondary oscillations will be overestimated (underestimated)  
by the coherent (totally incoherent)  approximation. In general all
we have said about peak position applies. The strength of the secondary
oscillations will fall somewhere in between the coherent and incoherent
predictions, the controlling factors being a combination of 
$\delta x_c$ and $\theta_c$.

For a given $x_c$ there will be a coherence time $\theta_c$ 
above which the perturbation is effectively coherent, 
and another $\theta_c$ below which the
totally incoherent approximation is a good approximation. The strength
of the secondary oscillations $\omega$ is a good indicator of how
good or bad the coherent (predicting $\omega=1$)
and totally  incoherent (predicting $\omega$ as in (\ref{epest}))
approximations are. We have found the rule that an
incoherent perturbation is effectively coherent if $\theta_c\gg
\sqrt{2}\delta x_c$, and effectively totally incoherent if
$\theta_c\ll \sqrt{2}\delta x_c$.

The situation for the defect defined in Sec.~\ref{gendef}
is shown in Figure~\ref{fig5}, where we show spectra in a grid
of values of $x_c$ and $\theta_c\approx 2.35\tau_c$. In the
second and third row we also present results in the totally incoherent
(dotted line) and coherent (dashed line) approximations. We see that 
large $x_c$ perturbations not only erase the secondary oscillations
if perfectly incoherent but also require unreasonably large
$\theta_c$ to deviate from the totally incoherent approximation. 
If $x_c=9$ (third column of Fig.~\ref{fig5}) then
only for $\theta_c\approx
20$ do any meaningful secondary oscillations exist. 
The coherent approximation requires
unacceptably large values for $\theta_c$. For all reasonable values
of $\theta_c$ the totally incoherent approximation, on the other hand,
simply smoothes out the already very soft or nonexistent
secondary oscillations. We will see that cosmic strings fall into this
category.

For perturbations with low $x_c$,
close to the causality lower bound,
smaller values of $\theta_c$ are required for effective coherence.
However these perturbations also 
show secondary oscillations in the totally incoherent approximation,
and therefore for all values of $\theta_c$
they do not discredit this approximation either.
The first column in Fig.~\ref{fig5} shows the case $x_c=3$. 
For $\theta_c<3$ the actual  oscillations are not far off the 
oscillations present in the totally incoherent approximation.
For $\theta_c\approx 6$ one already has $\omega\approx 0.9$,
and so the coherent approximation becomes a very  good quantitative
approximation. In this regime the real case and both approximations
do not differ significantly.

For intermediate values of $x_c$  there will 
normally be meaningful secondary oscillations but
peculiarly softer than coherent oscillations.
Both approximations clearly have shortcomings in this regime. 
In the second column in Fig.~\ref{fig5} we see what happens
for $x_c=6$ (texture type). One needs a rather large coherence time
($\theta_c\approx 20$ ) for the oscillations to be as
strong as coherent secondary oscillations (say $\omega=0.9$). 
On the other hand the totally 
incoherent approximation shows nearly non existent
secondary oscillations, a situation only realized
for $\theta_c<2$. For all the $\theta_c$ in between 
one should do the full calculation to get the right qualitative picture.

We may explain these facts semi-quantitatively by
repeating for incoherent perturbations the calculation
leading to the estimate (\ref{epest}). 
The structure function $x^{3/2}F^{1/2}(x)$ is then replaced by a step in $x$
centred at $x_c$ and with width $\delta x_c$, and ${\cal C}(x,x')$ 
by a step in $x-x'$ centred at zero and with width $\theta_c$.
By examining the third term in (\ref{pdelta}) one then sees that 
if $\theta_c>{\sqrt 2}\delta x_c$ the perturbation is effectively coherent
and $\omega=1$.  The analogue of (\ref{pdeltaap1}) for incoherent
perturbations is lengthy and unilluminating.
In the limit $\theta_c\ll {\sqrt 2} \delta x_c$ it may be 
approximated by the rectangular region of the integration domain: 
\begin{equation}\label{pdeltaap2}
k^3P(\Theta_0+\Psi)\approx{1\over 6}{\sin(\theta_c/(2{\sqrt 3}))\over
\theta_c/(2{\sqrt 3})}
{\left(
{\delta x_c-\theta_c/{\sqrt 2}\over \delta x_c}
-{\sin{\delta x_c-\theta_c/{\sqrt 2}\over {\sqrt 3}}\over  
\delta x_c/{\sqrt 3}}
\cos {2{x_*-x\over {\sqrt 3}}}\right)}
\end{equation}
and so
\begin{equation}\label{epest1}
\omega\approx{2\sin(\delta {\overline x}_c/{\sqrt 3})\over 
\delta {\overline x}_c/{\sqrt 3}+\sin(\delta {\overline x}_c/{\sqrt 3})}\; .
\end{equation}
with $\delta {\overline x}_c=\delta x_c-\theta_c/{\sqrt 2}$.
We see that finite coherence time tends to increase the oscillations
strength above their totally incoherent value, but not by much if
$\theta_c\ll {\sqrt 2} \delta x_c$.

\subsection{Translation into $C_l$'s}
\begin{figure}[htb]
\begin{center}
    \leavevmode
        {\hbox %
{\epsfxsize = 13cm\epsfysize=13cm
    \epsffile {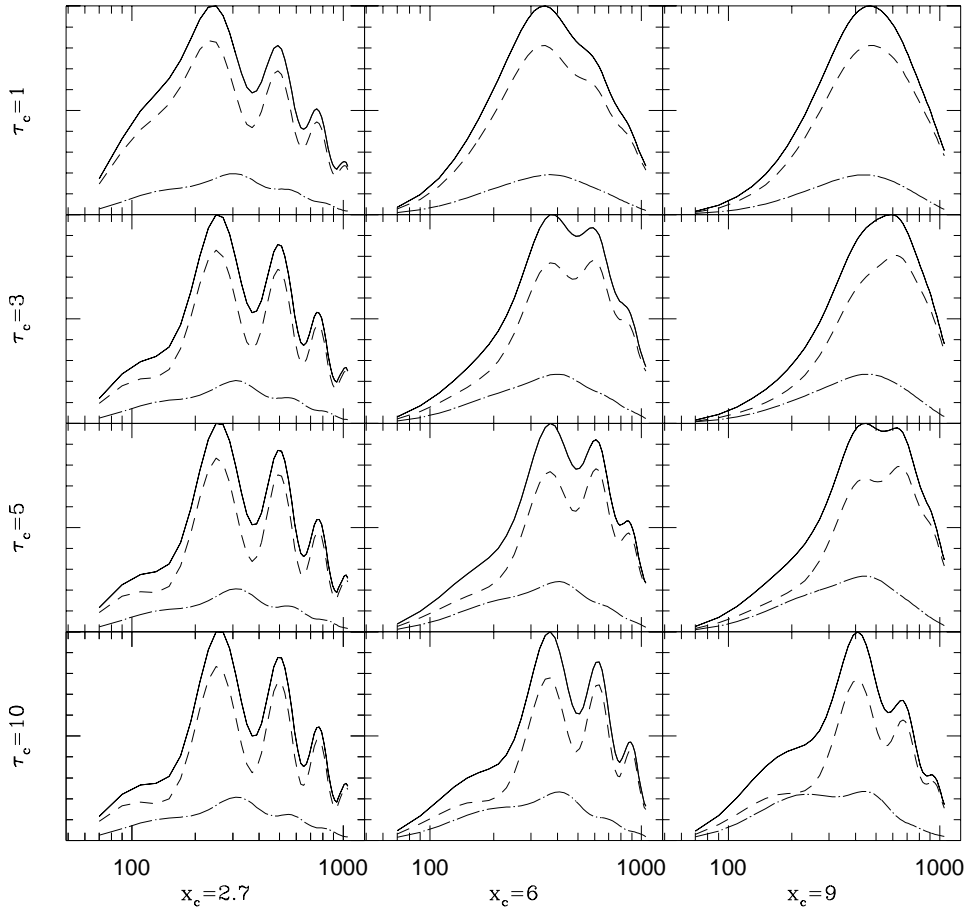} }}
\end{center}
\caption{$l(l+1)C_l$ spectra for a grid of models with various values
of $x_c$ (related to the defect coherence length) and 
$\theta_c\approx 2.35\tau_c$ (the defect coherence time). We have included
the monopole term (dash) and dipole term (point-dash), Silk damping,
and free-streaming. The monopole term is always dominant.}
\label{fig5a}
\end{figure}

With topological defects, as with inflation, the monopole power
spectrum is the dominant term in the Doppler peak region.
We  illustrate this statement in Fig.~\ref{fig5a}
using a grid of incoherent models as in Fig.~\ref{fig5},
with variable values for  $x_c$, and scaling
coherence time $\theta_c\approx 2.35\tau_c$.
We have included the monopole and dipole terms, but 
dropped the ISW term.
We also included Silk damping
and assumed that $\Omega_b=0.05$, $h=0.5$, and 
$T_0=2.726 K$ (where $T_0$ is the CMB average temperature).
We see that the dipole (Doppler) term is always subdominant,
and that the monopole free-streaming further softens the spectrum's
oscillations present in $P(\Theta_0+\Psi)$.

Figure~\ref{fig5a} confirms in terms of $C_l$'s what
we have inferred from the monopole power spectrum regarding
peak position and structure of secondary peaks. 
The salient features can be summarized as follows.
The $x_c$ parameter controls the peak position. For the
extreme value $x_c\approx 2.7$ the peaks appear on the 
adiabatic position. They suffer an additive shift to the
right for larger values of $x_c$. The strength of the secondary
oscillations depends on both $x_c$ (which is connected with structure
function width in realistic models) and $\theta_c$. For $x_c\approx
2.7$ there are secondary oscillations regardless of the exact
$\theta_c$ value. This is a confusing defect, as not only does it place
the Doppler peaks on the adiabatic position, but also the peak
structure is quite insensitive to the defect incoherence. 
For larger $x_c$ the secondary Doppler peaks survive only if 
the defect coherence time is much larger than $x_c$. This condition
seems unphysical for large $x_c$ so we expect realistic
defects with large $x_c$
not to have secondary oscillations.

Here is a rough guide to standard defect theories. Current
understanding places the
cosmic string models on 
the top right corner of Figure~\ref{fig5a} (large $x_c$, $\tau_c$
smaller than 3). They should have a single peak well after the main
adiabatic peak. Textures fall
somewhere in the middle of the figure ($x_c$ around 6, coherence
time not yet measured). Their main peak should be out of phase
with the adiabatic peaks. This is an accident related to the $x_c$
value for textures, and not a robust defect feature. Texture  secondary
oscillations should exist but be softer than predicted by the
coherent approximation (used in \cite{neil,durrer}). How 
much softer depends on the exact value of the texture's $\theta_c$.
If their coherence time is of the same order as strings ($\tau_c\approx
3$) their secondary oscillation will be very soft.

\section{Do cosmic strings have secondary oscillations?}\label{cs}
\begin{figure}[t]
\begin{center}
    \leavevmode
        {\hbox %
{\epsfxsize = 7cm\epsfysize=7cm
    \epsffile {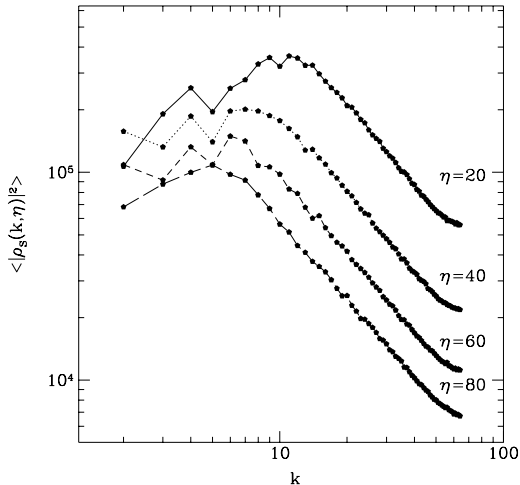} }}
\end{center}
\caption{The cosmic strings' energy
power spectra as measured from a simulation. This data was used to produce
the approximate fit to ${F_{\rho^s}(x)=P(\rho^s(k,\eta))\eta}$
given in (61).}
\label{fig6}
\end{figure}
Besides the general analysis performed in Section~\ref{genqual}
we may also target concrete defect scenarios.
The parameter space scanning methods developed in Section~\ref{genqual}
may then be useful in  allowing use of partial or uncertain
information obtained from simulations. The idea is to vary the physical
inputs of the calculation within the simulations'
uncertainties, or to fill in what simulations left undetermined
 in the most general form.
One may  then evaluate the impact of our uncertainties on the final result.
It may happen that in spite of simulation  uncertainties some
qualitative Doppler peak features are already  robust predictions.
We will argue that this is the case for the absence of secondary
oscillations in cosmic string scenarios.

\begin{figure}[t]
\begin{center}
    \leavevmode
        {\hbox %
{\epsfxsize = 6cm\epsfysize=6cm
    \epsffile {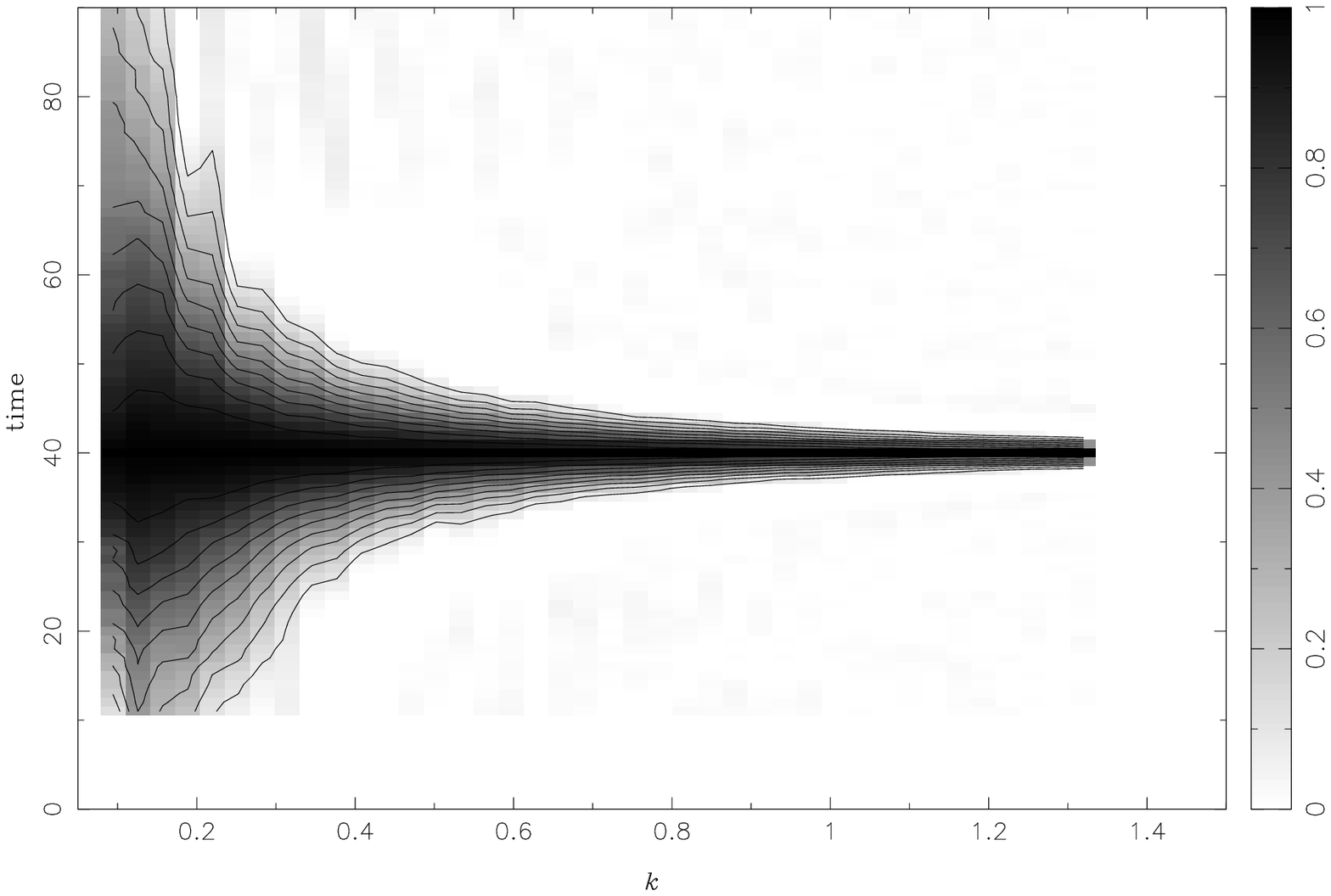} }
{\epsfxsize = 6cm\epsfysize=6cm
    \epsffile {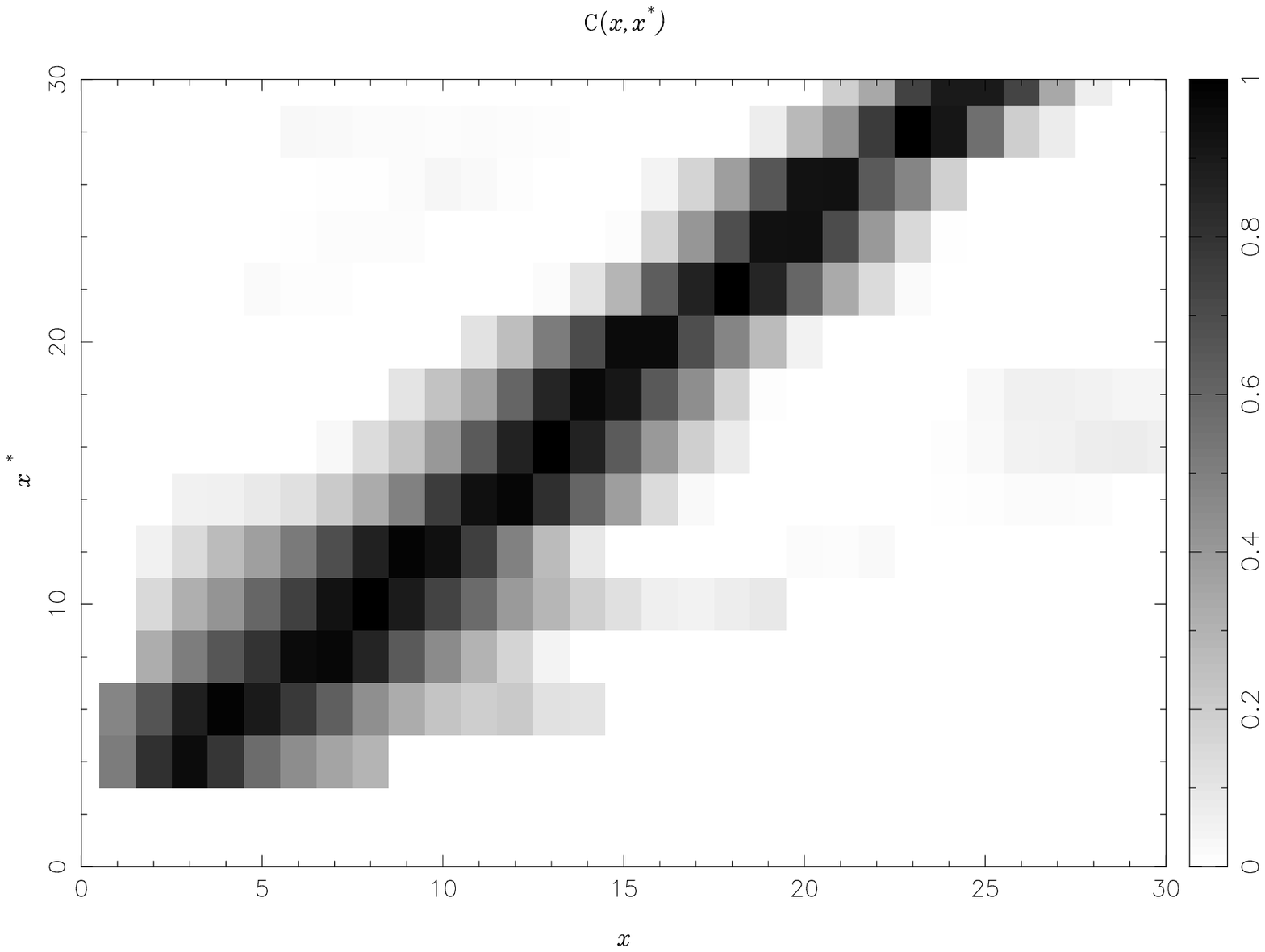} }}
\end{center}
\caption{The measured ${\rm cor}(\rho^s(k,40),\rho^s(k,\eta))$
and the coherence function ${\cal C}_{\rho^s}(x,x')$ it implies.
}
\label{fig7}
\end{figure}

\subsection{What can be measured from simulations}
In Fig.~\ref{fig6} we plot the structure function of the strings' energy
density defined by $P(\rho^s(k,\eta))={F^2_{\rho^s}(x)/\eta}$, as
discussed in \cite{realizations}. There we show that this
is well  fitted with
\begin{equation}\label{stenergy}
F_{\rho^s}^2=
{N\over (1-2bx_cx+bx^2)^{1/2}}
\end {equation}
where $N$ is a constant, $b\approx .006$, and $x_c\approx 10$.
Although we cannot measure the $x\ll 1$ region with the simulation
it is known that a white noise behaviour is to be expected. Therefore
we may safely take  (\ref{stenergy}) as a valid extrapolation.
The strings' energy coherence function
is plotted in Figure.~\ref{fig7}. On the right we plot
\begin{equation}
{\rm cor}(\rho^s(k,40),\rho^s(k,\eta))=
{{\langle \rho^s(k,40) \rho^s(k,\eta)\rangle}\over
\sigma(\rho^s(k,40))\sigma(\rho^s(k,\eta))}
\end {equation}
and on the left the ${\cal C}(x,x')$ inferred from this measurement.
The region $x<1,x'<3$ has been left undetermined by the simulation,
but since ${\cal C}(x,x')$ is symmetric this leaves undetermined
only $x<1,x'<1$.
We shall fill this gap with the widest range of possibilities. 
The last input required is $\Gamma$. In the Appendix we give some general
properties imposed on equations of state by energy conservation. 

\begin{figure}[t]
\begin{center}
    \leavevmode
        {\hbox %
{\epsfxsize = 12cm\epsfysize=12cm
    \epsffile {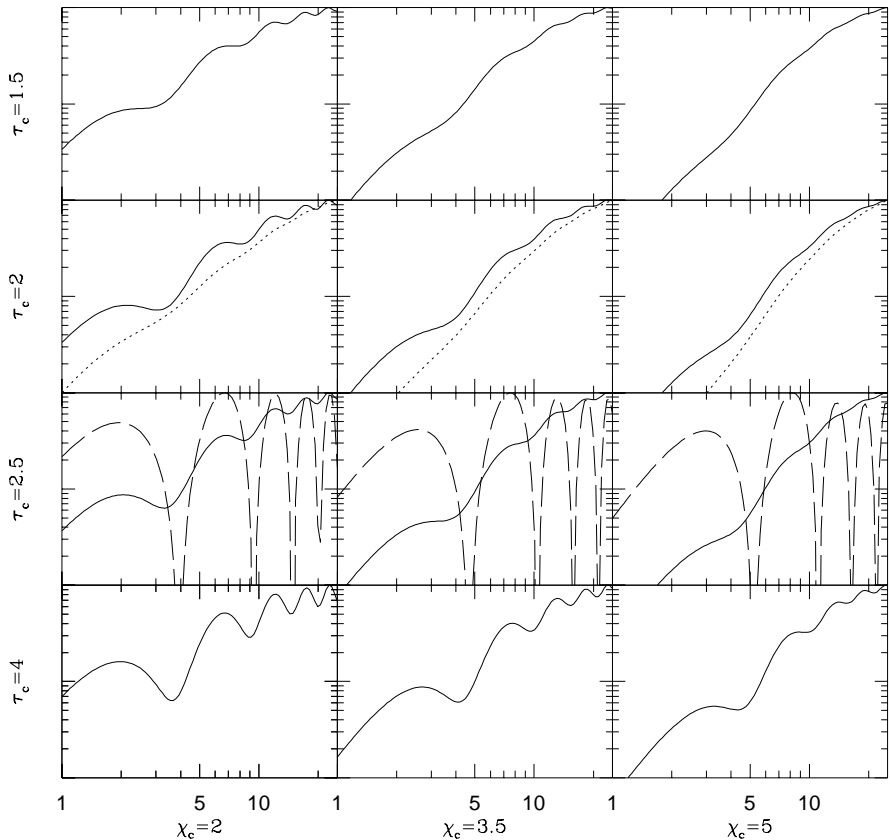} }}
\end{center}
\caption{The top three rows show a grid of cosmic strings'
spectra $P(\Theta_0+\Psi)$ for 
$\tau_c$ and $\chi_c$ varying within the allowed uncertainties.
In dashed and dotted lines we have plotted a coherent and a totally
incoherent approximation.
The last row addresses the issue of how big
the coherence time of a cosmic string would have to be for  secondary
oscillations with $\omega\approx 0.5$ to appear. The value $\tau_c=4$
is found, clearly ruled out by simulations.}
\label{fig8}
\end{figure}

\begin{figure}[t]
\begin{center}
    \leavevmode
        {\hbox %
{\epsfxsize = 9cm\epsfysize=9cm
    \epsffile {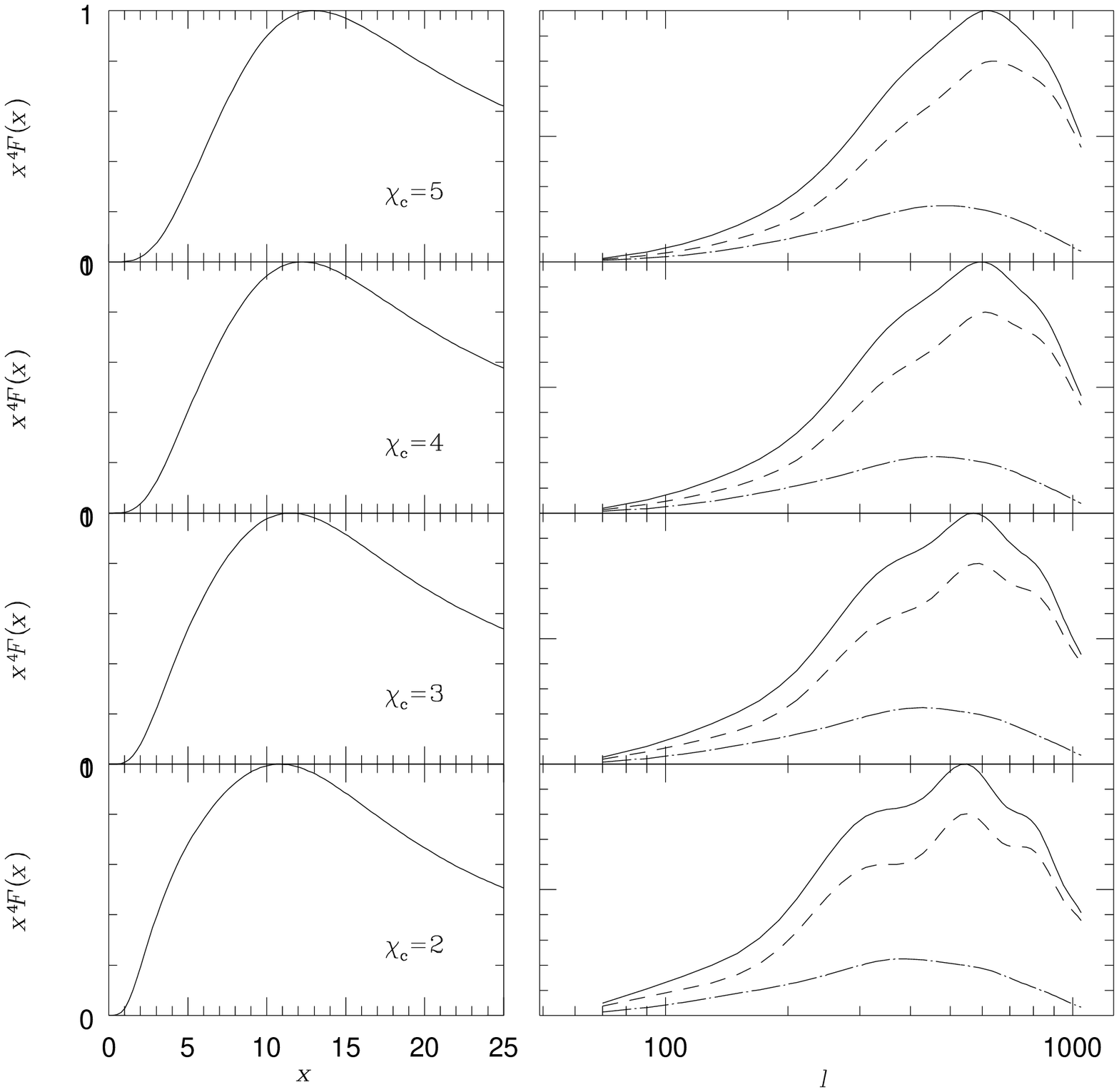} }}
\end{center}
\caption{The marginally acceptable (given simulations) 
cosmic string parameters which are most favourable for 
secondary oscillations: large coherence time
$\theta_c=6$, coherent filling in the region where ${\cal C}(x,x')$
could not be measured, low values of compensation scale $\chi_c$. 
On the right we show $l(l+1)C_l$
spectra from the monopole (dash) and dipole (point-dash) 
terms, in these cases. As for inflation the monopole dominates.
Even pushing causality ($\chi_c=2$) one does not obtain meaningful
oscillations.
On the left we show the structure functions $x^4F(x)$ for the various
$\chi_c$ parameters considered. Broad peaks centred around
$x=10$ are always obtained. }
\label{fig9}
\end{figure}

\subsection{The effects of the compensation and coherence time}
We assume that stress-energy components other
than $\rho^s$ can be ignored for the purpose of studying secondary
oscillations (we take this question up further in the appendix).
We therefore set 
$\Gamma=1$. The main uncertainty from the simulations concerns
the strings coherence time. From Figs.~\ref{fig7}
this can be estimated to be $\theta_c\approx 3-6$ (corresponding
to $\tau_c\approx 1.5-2.5$). There is some evidence that $\tau_c$
decreases at large $x$. We shall fill in the region $x<1,x'<1$
in three  ways:  same coherence function as in outer region;
incoherent filling, in which one sets $\tau_c=0$ in this region;
and coherent filling, in which one sets ${\cal C}=1$ in this region.
These cases cover the extreme possibilities for the behaviour
of   ${\cal C}$ where it was not measured.
Another uncertainty is the compensation
scale $\chi_c$. This may be  liberally placed  in the region
$2<\chi_c<5$. Within the framework of all that
is already known this uncertainty has little impact.

The top three rows of  
Fig.\ref{fig8} show a grid of spectra in which 
$\tau_c$ and $\chi_c$ are varied within the allowed uncertainties.
In dashed and dotted lines we have also plotted a coherent and a totally
incoherent approximation
(using $P$ rather than $P^r$).
We have redone Fig.\ref{fig8} with coherent, incoherent, and trivial
fillings in the $x<1,x'<1$ region and found the results nearly identical.
Clearly strings tend to erase secondary oscillations. 
For central values favoured by simulations this is done very 
effectively. For the marginally acceptable case where
$\tau_c=2.5$, $\chi=2$ we cannot rule out 
the existence of  very soft undulations in  $P(\Theta_0+\Psi)$.
The bottom row of Fig.\ref{fig8} addresses the issue of how big
the coherence time of a cosmic string would have to be for  secondary
oscillations to show up. Even accepting $\omega=0.5$ as a reasonable
oscillation strength, it appears that $\theta_c\approx 10$ is required.
This is ruled out by simulations.

In order to argue that  despite  simulation uncertainties,
we have established  cosmic string's  absence of secondary
oscillations, 
we now concentrate on the marginally acceptable scenario most favourable
for secondary oscillations. We choose the extreme value $\tau_c=2.5$
and fill the unmeasured domain of ${\cal C}$ coherently.
We then  allow $\chi_c$ to take values $2,\cdots, 5$.
As we have seen the monopole power spectrum then shows an
undulation. These are not even minima, but mere platforms in the rising
spectrum. In Fig.~\ref{fig9} we show the $C_l$ spectrum they translate 
into. We have included Silk damping, combined the monopole with the
dipole, and free streamed the result into $C_l$'s. 
Even in the most extreme case of
compensation scale $\chi_c=2$ (pushing causality) the resulting
$C_l$ undulations are extremely soft. Overall it is clear that 
the coherent approximation for cosmic strings will grossly
overestimate the oscillatory structure of the spectrum.
The totally incoherent approximation, on the other hand, 
simply exaggerates the lack of oscillations found in the real case.

This result may be understood by looking at the form of
the potential source term $x^4F(x)$ in Fig.~\ref{fig9}.
This  is dominated by the shape of
$F_{\rho^s}(x)$. A low compensation scale may enhance
the power on larger scales, and shift the peak in $x^4F(x)$ 
to the left. However
one would need to push causality limits in order to distort the strings'
peak significantly. $\chi=2$ is the most that can be done realistically.
In general $x^4F(x)$ for cosmic strings will have a 
very broad peak placed at $x=x_c\approx 10$. Hence each mode
will be active for a long time, requiring a very large coherence time
for effective coherence. The relatively low coherence time measured
then suggests effective total incoherence, and since $x^4F(x)$ is broad, this
will have time to erase the secondary oscillations. Also note that
the double integral (\ref{pdelta}) is dominated by a region centred
at $x=x'\approx 10$. Hence it is in this region that knowledge
of ${\cal C}(x,x')$ is most relevant. This explains why our results
are so insensitive to the way in which the unmeasured region $x<1,x'<1$
is filled. Since we also expect that the low $x$ region is where
corrections due to string loops and gravity waves are largest, the
unimportance of this region gives us added confidence in our
qualitative results.

\subsection{The cosmic strings  $C_l$ spectrum in the totally
incoherent approximation}\label{clscs}
\begin{figure}[t]
  \begin{center}
    \leavevmode
    {\hbox %
{\epsfxsize = 9cm\epsfysize=9cm
    \epsffile {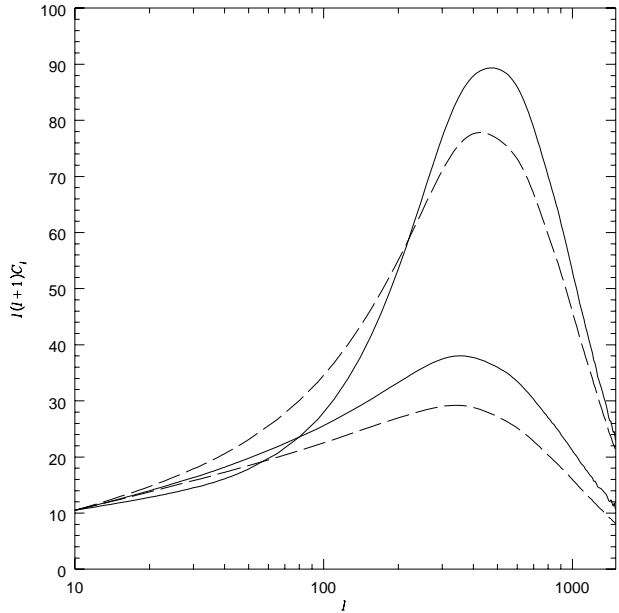} }}
    \caption{The $C^l$ spectrum for I (dash) and X (line) cosmic strings.
The top lines use $s=.1,\sigma=.4$ for X strings and
$s=.15,\sigma=.45$ for I strings. The bottom lines both use 
$s=.2,\sigma=.5$. We have assumed $\Omega=1$, $h=.5$, and
$\Omega_b=0.05$.}
    \label{fig11}
  \end{center}
\end{figure}
Since the totally incoherent approximation seems to be justified
for cosmic strings we have solved the full H+S algorithm in this
approximation. We use the expression from \cite{AS}
for the time-integrated power spectrum
\begin{equation}
  \label{prcs}
  P^r(\rho^s)={1\over 1+2(\beta x)^2}
\end{equation}
We consider the two cases 
$\beta=1$ and $\beta=.3$ similar to the X and I models in \cite{AS}.
We consider only scalar contributions. We assume that 
the defect variables 
are subject to equations of state of the form
$p^s=\gamma(x)\rho^s$, $\Pi^s=\eta^2\gamma_s(x)
\rho^s$, and $v^s=\eta\gamma_v(x)\rho^s$. Energy conservation
at small $x$ requires that $3\gamma(0)=(1/2\alpha)-1$ and 
$\gamma_v(0)=(1-2\alpha)/(3\alpha(4\alpha+1))$, with
$\alpha=\eta h$ (see appendix). We make use of a string
simulation to determine the large $x$ behaviour. We find,
with large uncertainties,  that
$x\gamma_v(x)= s\approx.1-.3$, and $x^2\gamma^s(x)=\sigma\approx.4-.55$. 
We interpolate between the $x\gg1$ and $x\ll 1$ behaviour.
We set $\Pi=0$ and assume that $a^2\rho\Delta_T$ is subdominant
except for the compensation and we  fix the compensation
scale at  $\chi_c=2\pi$. Using  (\ref{poteq1}) and (\ref{poteq2})
we finally obtain the required cosmic strings potential structure
functions to be inserted in the HS formalism as modified for
incoherent perturbations. The results are plotted in Fig.~\ref{fig11}.
The Sachs-Wolfe plateau exhibits a ``running'' tilt ranging from
$n\approx 1.4$ 
before $l=10$ to $n\approx 1.2$ at $30-40$, although this is quite
sensitive to input uncertainties. There is a single
Doppler bump located at $l\approx 400-600$.
These last two features are remarkably robust 
against uncertainties in the 
strings energy structure function, compensation  and equations of state. 
The peak position is affected mostly by the energy structure 
function.  The 
lower side of the $l\approx 400-600$ range appears to be favoured
by X strings, and the upper side by I strings. 
The effect of equations of state on the peak position is negligible. 
An unnaturally strong
compensation on small scales could push the peak further to the right
but this is unlikely. 
The absence of secondary oscillation is a permanent feature
whatever parameters changes one introduces, within
the ranges mentioned above. The ratio between 
the peak and the plateau heights, on the other hand, 
can change by as much as an order of magnitude. Even small changes
in the equations of state appear to be relevant. This is because 
the ISW plateau and the intrinsic terms probe different combinations
of defect stress-energy components. One may hope that the peak/plateau
height may act as a powerful probe of the defect speed and viscosity
composition. On the other hand a prediction of the peak/plateau
height ratio for strings  will have to wait for improved simulations.

\section{Conclusions}
We have  developed our formalism along two lines. We
isolated the differences between 
inflation and defects which we found important for Doppler peak features.
These were cast into the concepts of active and passive fluctuations,
and coherent and incoherent fluctuations. 
These concepts allowed us to discuss Doppler peak features for a generic 
abstract defect, before addressing any concrete example. In order to
address concrete examples
we had to develop formalism along a second line. We extended 
the Hu and Sugiyama formalism so as to accommodate topological 
defect theories. The extensions concern mainly the way in which
averages are taken when the photon fluid is being driven incoherently.
We also allowed for photon backreaction effects
so as to take into account the causality-required compensation.

We then derived two types of results. We studied Doppler peaks' position
and the structure of secondary oscillations for a generic defect, and then
for cosmic string theories. We found that generic defects place the primary
peak on or to the right of the adiabatic position ($l\approx 220$ for sCDM
cosmological parameter values). The shift to the right, when present,
is always additive, in contrast with low $\Omega$ models, which apply
a multiplicative type of shift. The value of the shift is controlled
by a single parameter $x_c$ which can be identified as  
$x_c \equiv 2\pi\eta/\xi_c$, where  $\xi_c$ is roughly the 
coherence length of the defect. Very large defects, on the verge of violating
causality, produce no shift. The smaller the defect coherence length,
the larger its shift. The texture out-of-phase signature found
by \cite{neil,durrer} is therefore an accident related to the particular
value of $x_c$ for textures, and not a generic defect feature.

We also found that the structure of secondary oscillations may be radically
different for generic defects. This is generally controlled by
the ratio of the defect scaling coherence time $\theta_c$ and $x_c$.
If $\theta_c\gg x_c$ then the defect is effectively coherent,
and displays secondary oscillations like the inflationary ones. 
If this is not the case then the secondary oscillations still
appear if the defect structure function is sufficiently narrow,
typically placing the primary peak close to the adiabatic position.
Defects with a main peak shifted to the right, however, typically have
broader structure functions. Provided that
$\theta_c$ is not larger than $x_c$ the secondary
oscillations get softer, the more so the further to the right the
main peak is. Anywhere to the right of the isocurvature position,
any defect with a realistic $\theta_c$ does not show secondary oscillations.

Given this picture there is good hope for a decisive experiment
confronting inflation and defects. Defects and inflation can only
be confused for $\Omega=1$ inflation, and a very large defect 
($x_c\approx 2.7$). Only this annoying defect would leave no
imprint of either its active or of its incoherent nature.
Any other defect would leave some exotic imprint on the peak
structure. An additive shift would ensure no confusion with low
$\Omega$ inflation. Also the secondary peaks would normally appear
as softer undulations  for not too large $x_c$, or totally
disappear for any realistic coherent times, if $x_c$ is large enough.
No inflationary scenario could realize these spectra. 

It remains to find out into which region of parameter space
($x_c$,$\theta_c$) each of the motivated defect scenarios
fall. This is clearly a quantitative issue to be decided from
simulations. We addressed this problem in connection with cosmic string
theories. We measured the required structure and coherence functions.
We found that even taking all the uncertainties into account
cosmic strings fall into the class of defect theories for which
the absence of secondary oscillations is a robust prediction.
This validates the totally incoherent approximation, and we solved
the H+S algorithm in this approximation for cosmic strings.
The solution reveals that given the simulation uncertainties
the main peak position should fall in $l\approx 400-600$.
The height of the peak, however, is more sensitive to simulation
uncertainties, and so we refrain from commenting on it at this stage.

\section*{Acknowledgements}
We acknowledge useful conversations with M. Hindmarsh, W. Hu,
N. Turok, M. White and we thank the many people at the
 Cambridge defects CMB meeting
last February who provided us with many insights and criticism.
J.M. thanks Kim Baskerville for all sorts of help in connection
with this project and
thanks St.John's College, Cambridge, for support. 
P.F. was supported by  the
Center for Particle Astrophysics, a NSF Science and
Technology Center at UC Berkeley, under Cooperative
Agreement No. AST 9120005.

\newpage

\section*{Appendix:
Active perturbations subject to equations of state}
\label{eqnst}
A complete solution for concrete defect models requires a precise
knowledge of the defect stress energy power spectra.  Knowledge of the
stress energy of other related matter components (such as the loops
and gravity
waves into which the cosmic strings decay) is also
essential\cite{caldwell}.  Furthermore, the correlations between all
these 
components must carefully taken into account.  In the case of textures
this problem is relatively simple because the textures and their
primary decay producs are all excitations of the same scalar field
which can be studied numerically.  The cosmic string problem is much
more complex, and further work is required before it is understood
clearly.  We present here the methodology used for this paper which
may, with further study, provide a sound basis on which to address
these questions.  Our starting point is the concept of defect
equations of state.

\subsection*{Coherent equations of state}
Active perturbations may be subject to at most two independent
coherent equations of state relating $\rho^s$, $p^s$, $v^s$, and $\Pi^s$.
Let them be  $p^s=\gamma(k\eta)\rho^s$, and $\Pi^s=\eta^2\gamma_s(k\eta)
\rho^s$ for a scaling active perturbation. 
For a coherent perturbation the conservation equations
will then impose another equation of state $v^s=\eta\gamma_v(k\eta)\rho^s$.
The equations of state impart perfect equal-time correlation between all 
four variables (${\rm cov}(A,B)=\sigma(A)\sigma(B)$
for any two variables) and so (\ref{2cori}) becomes the
more general statement
\begin{equation}\label{coheqns}
{\langle A({\bf k},\eta)B(-{\bf {k'}},\eta ')\rangle}=
{\langle A(-{\bf {k}},\eta)B({\bf k'},\eta ')\rangle}=
\delta({\bf k}-{\bf k'}) \sigma(A(k,\eta))
\sigma(B(k',\eta'))
\end{equation}
where $\sigma(.)={\sqrt{P(.)}}$.
This implies algebraic relations of the form $\langle
\rho^s v^s\rangle=\sigma(\rho)\sigma(v^s)$, $\langle
{\dot\rho}\rho\rangle={\dot\sigma}(\rho)\sigma(\rho)$, or
$\langle \dot\rho^2\rangle=({\dot\sigma}(\rho))^2$.
Using them one can square the conservation equations
(\ref{cons1}) and (\ref{cons2}), average, take the square root
of the result, and find that the variances $\sigma(\rho^s)$,
etc, satisfy the same linear conservations equation
as the variables $\rho^s$, etc, themselves. Introducing the variable
$x=k\eta$, writing $\sigma(\rho^s)=F_{\rho^s}(x)/\eta^{1/2}$ and
$\sigma(v^s)=F_{v^s}(x)\eta^{1/2}$ and denoting $d/dx$ by a prime
one has:
\begin{eqnarray}
x F'_{\rho^s}+ (\alpha(1+3\gamma)-1/2)F_{\rho^s}+x^2F_{v^s}
&=&0\label{cnssc1}\\
x F'_{v^s}+ {\left(2\alpha +{1\over 2}\right)}F_{v^s}+
{\left({2\over 3}x^2\gamma_s-\gamma\right)}F_{\rho^s}&=&0
\label{cnssc2}
\end{eqnarray}
where $\alpha=\eta{\dot a }/a$.
These can be solved expanding in Taylor series around $x=0$.
Causality implies that $F_{\rho^s}(0)=const$, and isotropy implies
that $\gamma_s(0)=const$. To zeroth order equations (\ref{cnssc1}) and
(\ref{cnssc2}) imply that 
\begin{eqnarray}
3\gamma(0)&=&{1\over 2\alpha}-1\label{ct1}\\
\gamma_v(0)&=&{1\over 3\alpha}{1-2\alpha\over 4\alpha+1}\; .
\label{ct2}
\end{eqnarray}
Fixing $\gamma_s(0)$ can also be done by extending the above
calculation up to second order. 

In this context the
gauge-invariant formalism has the advantadge over synchronous gauge
calculations that  
the potentials are also
subject to equations of state of the form
$\Phi=\eta^2\gamma_{\Phi}\rho^s$
and $\Psi=\eta^2\gamma_{\Psi}\rho^s$. This is because one may find a 
complete set of Einstein's equations which are elliptic. One must bear
in mind that the gauge invariant potentials may be expressed as
combinations of synchronous gauge variables, and their time derivatives. 
Hence it is possible
to get rid of potential time derivatives in Einstein's equations, assuming 
source conservation, and write equations of state for the potentials.
This advantage has one drawback: the defect structure
function appearing as a potential source now combines the defect energy
with its speed and viscosity, whereas in the synchronous gauge approach 
one only needs its pressure. Depending on the information available
on the defect 
this may or may not be a problem.
>From Einstein's
equations (\ref{poteq1}) and
(\ref{poteq2}) we have
\begin{eqnarray}
\gamma_{\Phi}&=&4\pi\gamma_c{1+3\alpha\gamma_v\over x^2}\label{gamaphi}\\
\gamma_{\Psi}&=&-4\pi{\left(\gamma_c{1+3\alpha\gamma_v\over x^2}
+2\gamma_s\right)}\; .\label{gamapsi}
\end{eqnarray}
These allow us to write the required relations
\begin{eqnarray}
\sigma(\Phi G)=&\eta^{3/2}\gamma_{int}F_{\rho^s}=&
\eta^{3/2}{\gamma_{\Phi}F_{\rho^s}\over(1+R)^{1/4}}
{\left(1-(1+R){\gamma_{\Psi}\over \gamma_{\Phi}}
+{3{\ddot R}\over 4k^2}-J^2\right)}
\end{eqnarray}
Using the conservation equations (\ref{cons1}) and (\ref{cons2})
we could also write $\dot\Phi$ and $\dot\Psi$ in terms of defect
variables (and not their derivatives) and find a similar equation
of state for the potentials appearing in the ISW term. The result
is lengthy but straightforward to derive.

\subsection*{Small scale equations of state}
On small scales the defects cannot be subject to
coherent equations of state.
For realistic sources it happens that $F_{\rho^s}(x)\propto
x^{-n}$, and $\gamma$ and $k^2\Pi^s/\rho^s=x^2\gamma_s$ tend 
to a constant for $x\gg 1$. One may check 
that equations (\ref{cnssc1}) and (\ref{cnssc2})
would then lead to over-restrictive constraints (like $n=
2\alpha-3/2$). We find from simulations that rather than
relations like (\ref{coheqns}) on small scales one
has
\begin{equation}
  {\rm cor}( A({\bf k},\eta),B(-{\bf k}',\eta'))\approx 0
\end{equation}
where $A$ and $B$ are any two defect stress-energy components.
This means that the different defect stress-energy components are
in fact uncorrelated random variables, for which it makes more sense
to postulate equations of state of the form:
\begin{eqnarray}
  {\langle p^{s2}\rangle}&=&\gamma^2{\langle \rho^{s2}\rangle}\nonumber\\
    {\langle v^{s2}\rangle}&=&\eta^2
         \gamma_v^2{\langle \rho^{s2}\rangle}\nonumber\\
     {\langle \Pi^{s2}\rangle}&=&\eta^4
         \gamma_s^2{\langle \rho^{s2}\rangle}    
\end{eqnarray}
We have found for cosmic strings that for large $x$
\begin{eqnarray}
  x\gamma_v(x)&=& s\approx.1-.3\nonumber\\
  x^2\gamma^s(x)&=&\sigma\approx.4-.55
\end{eqnarray}

\subsection*{The effect of the equations of state on cosmic strings
coherence properties}
>From what we have said above the low $x$ behaviour of the $\Gamma$
factor (defined in (\ref{gamadef})) is 
\begin{equation}
  \Gamma^2={\left(1+{1-2\alpha\over 1+4\alpha}+\gamma_s(0)\right)}^2
\end{equation}
whereas its high $x$ behaviour is of the form
\begin{equation}
  \Gamma^2=1+9\alpha^2{s^2\over x^2}+\sigma^2
\end{equation}
In both regimes the $\Gamma$ factor is of order 1.
Therefore $\Gamma$ may 
never modify the shape of the structure function $x^4F(x)$
as given by (\ref{gamadef}), which is determined by $F_{\rho^s}$ and
the compensation factor. For this reason we simply set $\Gamma=1$ in
our discussion of strings coherence features. On the other hand no doubt
$\Gamma$ will affect the relative normalization of the structure functions
used in the  intrinsic and the ISW terms, which use different equations
of state. We have found the the relative height of the peak and low $l$
plateau reflect sensitively the defect speed and viscosity composition.

\subsection*{Incoherent perturbations}
If the perturbation is not coherent but is still subject 
to two equations of state,
then all we have said still holds for the power spectra of the
perturbation, but not for its
time-integrated power spectra. One may try to relate power spectra
and time-integrated power spectra. Suppose that one can factorize
the unequal-time correlator as
\begin{equation}
P(\rho^s(k,\eta),\Delta\eta)=\sigma(\rho^s(k,\eta))\sigma
(\rho^s(k,\eta+\Delta\eta)){\cal C}(k,\Delta \eta)
\end{equation}
for, say, $\rho^s$. ${\cal C}(k,\Delta \eta)$ is a coherence function
not necessarily symmetric about zero in $\Delta\eta$. Then wherever ${\cal C}
\approx 1$ there is coherence, which is then cut off as ${\cal C}$
goes to zero for $\Delta\eta>\tau_c$. Suppose that this transition
is very abrupt (e.g. ${\cal C}$ is nearly a step function).
Then we can expand the time-integrated power spectrum written as
in (\ref{prpk}) in the form
\begin{equation}
P_r(\rho^s(k,\eta))=\tau_c^{(1)} P(\rho^s(k,\eta))+\tau_c^{(2)}
\sigma(\rho^s(k,\eta)){\dot\sigma}(\rho^s(k,\eta))+{\tau_c^{(3)}
\over 2}\sigma(\rho^s(k,\eta)){\ddot\sigma}(\rho^s(k,\eta))+
{\cal O}(\tau_c^4)
\end{equation}
where
\begin{equation}
\tau_c^{(n)}={{\int}^{\infty}_{-\infty}}\,d\Delta\eta\;
{\cal C}(k,\Delta \eta)(\Delta\eta)^{n-1}\; .
\end{equation}
We have $\tau_c^{n}
={\cal O}(\tau_c^n)$. However, if  ${\cal C}$ is symmetric
about zero all the even $n$ vanish. We shall assume that this is not
the case, so that the term in $\tau_c^{(2)} $ is the
dominant correction to order ${\cal O}(\tau_c^2)$. Using the
conservation equations we can then write 
\begin{equation}\label{eqn1}
P_r(\rho^s(k,\eta))=\tau_c P(\rho^s(k,\eta))
{\left( 1-{\hat\tau_c^{(2)} }(\alpha(1+3\gamma)+x^2\gamma_v)
\right)}
\end{equation}
with ${\hat\tau_c^{(2)} }=\tau_c^{(2)}/(\tau_c\eta)$. 
For high $k$ the behaviour of $P$ and $P_r$ may be very different.
For instance, for cosmic strings it is known that 
$P_r(\rho^s(k,\eta))\propto k^{-2}$ whereas 
$P(\rho^s(k,\eta))\propto k^{-1}$ (\cite{AS}).
However for small $k$ equation (\ref{eqn1}) tells us that
that $P_r$ and $P$ are related simply
by a multiplicative constant. It can be checked that this is
true for the power spectra of any variable.
Moreover the multiplicative constant is the same ($\tau_c^{(1)}$)
for all variables if one ignores corrections in ${\hat\tau_c^{(2)} }$.
Therefore to this level of approximation the large scale behaviour
of $P_r$ can be simply inferred from (\ref{ct1}) and (\ref{ct2}).
If, however,  one is to keep corrections of order 
${\hat\tau_c^{(2)} }$ then the situation is more complicated, 
as the multiplicative constant depends on the variable.
For a quantity related to $\rho^s$ by an equation of state of the form
$\Gamma(k\eta)\rho^s$ one then has for $x\rightarrow 0$ 
\begin{equation}
P_r(\Gamma(k\eta)\rho^s(k,\eta))=\tau_c \Gamma^2(k\eta) P(\rho^s(k,\eta))
{\left( 1-{\hat\tau_c^{(2)} }(\alpha(1+3\gamma)
+\Gamma'/\Gamma) \right)}\; .
\end{equation}
Higher derivatives of the $\gamma$'s at zero would 
be required to ${\cal O}({\hat\tau_c^{(2)} })$.  How important
these corrections are remains to be determined.

Modelling the $P_r$ on small scales is again more complicated and defect
dependent. If the $\gamma$'s are always slowly varying one may
use an approximation where formally $P_r=\eta\tau_c^s\sigma^{r2}$,
where $\tau_c^s=\tau_c/\eta$, and $\sigma^r$ means the value of
$\sigma$ as computed for coherent fluctuations but with $P(\rho^s)$ 
replaced by $P_r(\rho^s)$. Then the required potential properties
can be found from
\begin{eqnarray}
P_r(\Phi G)&=&\tau_c^s \eta^4\gamma_{int}^2F^{r2}_{\rho^s}\\
P_r({\dot\Psi}-{\dot\Phi})&=&\tau_c^s \eta^2\gamma_{isw}^2
F^{r2}_{\rho^s}
\end{eqnarray}
with $\gamma_{int}$ and $\gamma_{isw}$ are the equations of state for
the potentials appearing in the intrinsic and ISW terms.
We have used this approximation in the calculation in Section~\ref{clscs}
of the $C_l$ spectrum of cosmic strings in the totally incoherent 
approximation.


\begin{thebibliography}{99}
\bibitem{exp}
M. White, D. Scott and J. Silk, 
{\it Annu. Rev. Atron. Astrophys.}{\bf 32} 319-370 (1994).
\bibitem{infl}
P. Steinhardt, {\it Cosmology at the
Crossroads} to appear in the {\it Proceedings of  the Snowmass Workshop on
Particle Astrophysics and Cosmology}, E. Kolb and R.Peccei,
eds. (1995)   astro-ph/9502024.
\bibitem{defc}
T.W.B. Kibble, {\it J. Phys.}, {\bf A9} 1387-1398 (1976),
A. Vilenkin and P. Shellard, {\it Cosmic
Strings and other Topological Defects.} Cambridge University Press,
Cambridge (1994). M. Hindmarsh and T.W.B. Kibble, Reports on Progress in
Physics, 58,  477-562 (1995) 
\bibitem{HS}
W.Hu and N.Sugyiama, {\it Astroph.J.}, {\bf 444} 489 (1995),
W.Hu and N.Sugyiama, {\it Phys. Rev. D}, {\bf 51} 2599-2630 (1995).
\bibitem{battye} R. Battye,  Small-angle anisotropies in the CMBR from
active perturbations.  
IMPERIAL/TP/95-96/46 
\bibitem{BE}
Bond J. R., Efstathiou G., {\it MNRAS} {\bf 226} 655 (1987).
\bibitem{crit}
G. Jungman, M. Kamionkowski, A. Kosowsky and D. Spergel, 
{\it Phys. Rev. Lett.} {\bf 76} 1007 (1996).
\bibitem{bsb} David Bennett, Albert Stebbins, and Fran\c cois Bouchet {\sl The
Astrophysical Journal, {\it Letters}}, {\bf 399}, L5-8. (1992)
\bibitem{b&r} D. Bennett and S. Rhie,
ApJ 406, L7 (1993).
\bibitem{reion}D. Coulson, P. Ferreira, P. Graham and N. Turok,
{\it Nature} {\bf 368}  27-31 (1994).
\bibitem{neil}
R. Crittenden and N. Turok, {\it Phys. Rev. Lett.}, {\bf 75}
2642 (1995).
\bibitem{durrer}R. Durrer, A. Gangui and M. Sakellariadou, 
{\it Phys. Rev. Lett.}, {\bf 76} 579 (1996).
\bibitem{us}
A. Albrecht, D. Coulson, P. Ferreira, and J. Magueijo,
{\it Phys. Rev. Lett.} {\bf 76} 1413-1416 (1996).
\bibitem{trasch12}
J.Traschen, {\it Phys. Rev. D}, {\bf 31} 283-289 (1985); 
J.Traschen, {\it Phys. Rev. D}, {\bf 29}  1563-1574 (1984).
\bibitem{james}J. Robinson and B. Wandelt, 
{\it Phys.Rev.D} {\bf 53} 618 (1996).
\bibitem{causal-ngt} N. Turok, Causality and the doppler
peaks. astro-ph/9604172/
\bibitem{steb}
S. Veeraraghavan and A. Stebbins, {\it Astrophys.J.}
{\bf 365} 37-65 (1990).
\bibitem{pen}U. Pen, D.N. Spergel and N. Turok, {\it Phys. Rev. D},
{\bf 49} 692-729 (1994).
\bibitem{usagain}
J.Magueijo, A. Albrecht, D. Coulson, P. Ferreira, {\it Phys. Rev. Lett},
{\bf 76} 2617 (1996).
\bibitem{traschk4}
L.F.Abbott and J.Traschen, {\it Astrophys.J.} {\bf 302} 39-42 (1986).
\bibitem{KS}
H.Kodama and M.Sasaki, {\it Prog.Th.Phys. Supp.}, {\bf 78} 1-166 (1984).
\bibitem{mypaper}
J.Magueijo, {\it Phys. Rev. D}, {\bf 46} 3360-3371 (1992).
\bibitem{causal}
J.Magueijo et al, 
Causal compensated perturbations in the CMBR, in preparation.
\bibitem{AS}
A.Albrecht and A.Stebbins,{\it Phys. Rev.
Lett.} {\bf 68} (1992) 2121-2124; {\it ibid.} {\bf 69}
2615-2618 (1992).
\bibitem{realizations}
P.Ferreira, A.Albrecht, D.Coulson, and J.Magueijo, in preparation.
\bibitem{HW}
Hu and White, Acoustic signatures in the cosmic microwave background,
IASSNS-96/6.
\bibitem{definfl}
J.Magueijo et al, Confusing defects and inflation, in preparation.
\bibitem{caldwell}  P. Avelino and  R. Caldwell,
Entropy perturbations due to cosmic strings astro-ph/9602116 (1996).
\end{thebibliography}
\end{document}